\documentclass[a4paper,12pt]{article}
\usepackage{geometry}
\usepackage[T1]{fontenc}
\usepackage{lmodern}
\usepackage{microtype}
\usepackage{amsmath,amssymb,bm}
\providecommand{\ket}[1]{\left|#1\right\rangle}
\providecommand{\bra}[1]{\langle#1|}
\providecommand{\braket}[2]{\langle#1|#2\rangle}

\providecommand{\Hhat}{\hat{H}}

\providecommand{\crea}[1]{\hat{a}^{\dagger}_{#1}}
\providecommand{\anni}[1]{\hat{a}_{#1}}

\providecommand{\occ}{\mathcal{O}}

\providecommand{\mHa}{\,\mathrm{mHa}}

\usepackage{graphicx}
\usepackage{booktabs}
\usepackage{caption}
\usepackage{subcaption}
\usepackage[numbers,sort&compress]{natbib}
\usepackage[colorlinks=true,linkcolor=blue,citecolor=blue,urlcolor=blue]{hyperref}
\title{\textbf{Rotated-Frame Determinant Discovery for Ground- and Excited-State Quantum Selected Configuration Interaction}}
\author{Yukihiro Okuno \\ FUJIFILM Corporation, Minamiashigara, Kanagawa 250-0193, Japan}
\date{\today}
\begin{document}
\maketitle
% ======================
\begin{abstract}
  Quantum Selected Configuration Interaction (QSCI) constructs a variational determinant space from measurements of a
  prepared electronic state, but finite-shot sampling repeatedly observes high-probability determinants while missing
  configurations needed for an accurate selected-CI description.
  We introduce rotate--sample--unrotate (RSU) determinant discovery, in which the same prepared state is sampled in
  complementary orbital frames and the observed rotated determinants are used to generate a finite set of candidate
  determinants in the original canonical basis. The resulting Hamiltonian is diagonalized in an orthonormal
  selected-CI space,
  without enumerating the full determinant space or retaining a nonorthogonal final basis.
  We construct state-targeted orbital frames from an initial selected-CI wavefunction and its external residual to
  enhance the recovery of determinants relevant to the target electronic state. We further combine RSU with the
  generative quantum eigensolver (GQE) to prepare sampling states with enhanced target-state character, particularly for
  excited states that may be poorly represented by ground-state-oriented sampling. For excited states, candidate GQE
  circuits are evaluated by classical selected-CI diagonalization in finite sampled determinant spaces,
  so that the quantum
  procedure requires only state preparation and determinant sampling.
  For N$_2$, exact-state controls isolate the determinant-discovery effect: at the stretched geometry,
  residual-natural-orbital RSU reaches chemical accuracy with 5,000 shots, compared with 20,000 shots
  for conventional QSCI. The improvement persists with approximate GQE-prepared states and is also observed
  for target excited states. These results establish rotated-frame sampling as a route to improving
  finite-shot determinant discovery in QSCI for both ground and excited states.
\end{abstract}

% ======================================================================
\section{Introduction}\label{sec:intro}
% ======================================================================
\indent Quantum computers offer a potential route to electronic-structure calculations
that become prohibitively expensive for exact classical methods as the number
of electrons and orbitals increases \cite{McArdle2020,Cao2019,Bauer2020}. In the fault-tolerant regime,
quantum phase estimation provides a route to molecular energy eigenvalues~\cite{Aspuru2005,Abrams1999},
but its circuit-depth
and coherence requirements remain beyond the capabilities of present-day hardware \cite{McArdle2020,Cao2019}.
For noisy intermediate-scale quantum (NISQ) devices, the variational quantum eigensolver (VQE) has
therefore been studied extensively \cite{Preskill2018,Peruzzo2014,McClean2016}. 
Its practical application to chemically accurate calculations remains challenging, however,
because energy estimation requires many measurements and circuit optimization can be hindered by difficult
optimization landscapes, including barren plateaus \cite{mcclean2018barren,Gonthier2022,okuno2025}.

\indent
These difficulties motivate a different division of labor between quantum
and classical computation. Rather than estimating the energy through
repeated quantum measurements, a quantum device can help identify a useful
subspace in which the electronic Hamiltonian is subsequently diagonalized
classically.
This perspective connects naturally to selected configuration interaction (SCI) \cite{harrison1991}.
Classical SCI methods approximate the full configuration interaction (FCI) wavefunction by retaining
a carefully chosen subset of Slater determinants. For example, CIPSI uses perturbative estimates to guide
determinant selection \cite{Huron1973}, whereas heat-bath configuration interaction (HCI) screens
candidate determinants using their Hamiltonian couplings to the current wavefunction \cite{Holmes2016}.

\indent Quantum-selected configuration interaction (QSCI) retains the selected-subspace framework but uses a
quantum device to propose determinants \cite{kanno2023qsci}. An approximate electronic state is prepared and
measured in the computational basis, yielding bitstrings that correspond to candidate Slater determinants.
The Hamiltonian is then constructed and diagonalized classically in the resulting subspace.
Sample-based quantum diagonalization (SQD) follows this general strategy and incorporates
post-processing of noisy measurement outcomes \cite{robledo2024sqd}.
Classical determinant-space expansion can also be applied after quantum
sampling. For example, Ext-SQD augments a sampled determinant space using
single and double excitations before classical diagonalization~\cite{barison2025quantum}.
QSCI has also been applied to excited-state
calculations~\cite{Shirai2025,Sugisaki2025HSBQSCI}.

\indent This QSCI approach has progressed beyond small proof-of-principle calculations. SQD has been demonstrated
on superconducting hardware for molecular systems including N$_2$ and iron--sulfur clusters,
using circuits of up to 77 qubits \cite{robledo2024sqd}. Hamiltonian simulation-based QSCI has also
been demonstrated on quantum hardware for carbon-chain molecules using up to 36 qubits \cite{Sugisaki2025HSBQSCI}.
These studies establish the feasibility of sampling electronic configurations on devices with tens of qubits,
although the accuracy and computational value of the resulting selected subspace depend on which
configurations are actually found.

\indent Despite these hardware demonstrations, efficient quantum sampling remains a central challenge for QSCI.
Although classical diagonalization avoids the statistical uncertainty associated with measuring
the Hamiltonian expectation value, the selected determinant set still depends on
stochastic measurement outcomes \cite{Reinholdt2025}. Reinholdt et al. showed that repeated
sampling of already observed determinants can make the discovery of new configurations increasingly inefficient,
even when sampling from an exact ground-state wavefunction \cite{Reinholdt2025}.
Conversely, a broader sampling distribution can reveal more distinct determinants but may produce a
less compact CI expansion, increasing the cost of classical diagonalization at a given
accuracy. Sampling from an approximate prepared state introduces a further difficulty
if its determinant probabilities do not favor configurations needed for an accurate
subspace. Thus, the relevant objective is not simply to increase the number of
distinct measurement outcomes, but to identify useful determinants within both a limited number
of measurements and a manageable diagonalization subspace.

\indent A further limitation arises when a single prepared state $\ket{\Psi}$ primarily
approximates the ground state. Its measurement distribution then favors ground-state configurations,
while determinants important to excited states may be sampled too rarely to enter the selected subspace.
This can leave the excited-state energies overestimated and motivates state-aware preparation and
sampling for multiple target states.

\indent We address these challenges by separating two questions that are often treated together in QSCI:
how to prepare states relevant to the target electronic levels, and how to extract useful determinants
from those states with a finite number of measurements. For the latter, we exploit the freedom to change
the orbital basis in which a state is measured. A rotated determinant that is readily observed can provide
candidates that are rarely sampled directly in the canonical orbital
basis. We therefore propose rotate--sample--unrotate (RSU) discovery:
the prepared state is sampled in different orbital frames, and the observed
rotated determinants are used to generate a finite set of canonical
candidates. The selected Hamiltonian is then diagonalized classically in
the resulting orthonormal canonical determinant space, without requiring
a nonorthogonal final eigenvalue problem.

\indent The effectiveness of determinant discovery also depends on the
prepared state. We use the generative quantum eigensolver
(GQE)~\cite{nakaji2024gqe} to construct separate state-preparation circuits
for the ground state and the target excited state, rather than relying
on samples from a single ground-state-oriented reference. GQE trains a
classical generative model to propose ordered excitation-operator
sequences, which define candidate quantum circuits and are evaluated
using the corresponding objective values. Unlike gradient-based VQE
optimization, this procedure does not require gradients of
quantum-circuit parameters.

For the target excited state, the key question is how to evaluate candidate
circuits so that they provide useful determinants for the subsequent
selected-CI calculation. Previous excited-state QSCI calculations have
used target-state approximations prepared by variational quantum deflation
with spin and overlap penalties~\cite{Shirai2025}, or Hamiltonian time
evolution from suitable initial states to generate the sampled determinant
spaces~\cite{Sugisaki2025HSBQSCI}.
In the present approach, each candidate excited-state GQE circuit is
evaluated using the target-state energy obtained from a classical
selected-CI calculation. The sampled determinants define a finite,
spin-completed space in which the lowest singlet energy is determined
subject to the target spatial symmetry and orthogonality to a fixed
approximate ground state. This energy is used to train the generative
model and guide the generation of subsequent candidate circuits.
The circuit-optimization objective is therefore the target-state energy
obtained in the sampled selected-CI space, rather than the Hamiltonian
expectation value of the prepared quantum state. Energy, spin, and
fixed-ground orthogonality are treated classically within this finite
space. The quantum evaluation of each candidate circuit requires only
state preparation and determinant sampling, without measurements of the
Hamiltonian, $\hat S^2$, or the ground-state overlap.
The resulting ground- and excited-state circuits supply state-specific
sampling distributions for QSCI and RSU. State preparation determines
the distribution being sampled, whereas RSU improves the discovery of
useful canonical determinants through sampling in different orbital
frames. The selected determinants are then used for classical
Hamiltonian diagonalization in a finite canonical selected-CI space.
Although GQE is used here, RSU can be applied independently of the method
used to prepare the sampling state.

\indent The remainder of the paper is organized as follows.
Section~\ref{sec:theory} develops the theoretical framework: the
reference-state provider, rotated-frame (RSU) discovery with TC-NO, SCI-NO,
and RES-NO frames, and orthogonal diagonalization for the ground and excited
states (Fig.~\ref{fig:pipeline-text}). Section~\ref{sec:comp} summarizes the computational
details, including the quantum/classical division of labor.
Section~\ref{sec:results} presents the ground- and excited-state
calculations, followed by a discussion of the results and limitations in
Sec.~\ref{sec:discussion}.

% ======================================================================
\section{Theoretical Framework}
\label{sec:theory}
% ======================================================================
\subsection{Rotate–sample–unrotate (RSU) determinant discovery}
\label{sec:stage3-discovery}
We introduce rotate--sample--unrotate (RSU) discovery, which uses measurements
in multiple orbital frames to propose canonical determinants for selected CI.
We describe the frame-dependent sampling distributions, identification of canonical determinants, and construction of the orbital frames.
\paragraph{Sampling distributions in orbital frames.}
\indent Let $n_{\mathrm{orb}}$ be the number of active spatial orbitals and let
$N_{\alpha}$ and $N_{\beta}$ be the fixed electron numbers in the two spin
sectors. The full particle-number-conserving canonical determinant sector is
\begin{equation}
\mathcal{X}=\{x=(\mathcal{O}_{\alpha},\mathcal{O}_{\beta}):
|\mathcal{O}_{\alpha}|=N_{\alpha},
|\mathcal{O}_{\beta}|=N_{\beta}\},
\qquad
|\mathcal{X}|=\binom{n_{\mathrm{orb}}}{N_{\alpha}}
\binom{n_{\mathrm{orb}}}{N_{\beta}}.
\label{eq:rsu-sector}
\end{equation}
Here $\mathcal{O}_{\sigma}$ is the set of occupied active orbitals of spin
$\sigma$ in the bitstring $x$. For a closed-shell sector,
$N_{\alpha}=N_{\beta}=N_{\mathrm{occ}}$.
Let $\ket{\Psi_k}$ be a prepared reference for target state $k$.
Conventional QSCI measures this state in a fixed computational basis and samples
a determinant $x\in\mathcal{X}$ with probability
\begin{equation}
  P_{k,I}(x)=|\braket{x}{\Psi_k}|^2.
  \label{eq:canonical-born}
\end{equation}
Here, $I$ denotes the identity orbital frame, corresponding to measurement in the canonical orbital basis
without rotation.
A determinant that contributes to an accurate selected-CI description may nevertheless have a low measurement probability in the canonical orbital basis and may therefore be missed with a finite number of shots. Measurements in rotated orbital frames can expose such determinants more efficiently.
We use an orbital frame $F=(R_\alpha,R_\beta)$, where $R_\alpha$ and $R_\beta$ are
orthogonal rotations of the spatial orbitals for the two spin sectors.
The corresponding fermionic Gaussian unitary implements a Thouless orbital rotation,
\begin{equation}
  \hat U_F = e^{\hat Z_F}, \qquad \hat Z_F = \sum_{\sigma\in\{\alpha,\beta\}} \sum_{pq} (\kappa_\sigma)_{pq} \hat a^\dagger_{p\sigma}\hat a_{q\sigma}, \qquad \kappa_\sigma^\dagger=-\kappa_\sigma, \qquad R_\sigma=e^{\kappa_\sigma}.
\end{equation}
Here $p$ and $q$ label active spatial orbitals. The anti-Hermitian one-body generator rotates
the creation operators according to
\begin{equation}
  \hat U_F\hat a^\dagger_{q\sigma}\hat U_F^\dagger = \sum_p (R_\sigma)_{pq}\hat a^\dagger_{p\sigma}.
\end{equation}
Let $|d\rangle$ denote a computational-basis determinant.
Consequently, $\hat U_F|d\rangle$ remains a single Slater determinant, formed from the
rotated occupied orbitals. In RSU, this transformation defines the measurement frame. 
A computational-basis measurement after applying $\hat U_F^\dagger$ to the reference samples
\begin{equation} P_{k,F}(d) =|\bra{d}\hat U_F^\dagger\ket{\Psi_k}|^2 =|\braket{\varphi_{F,d}}{\Psi_k}|^2, \qquad \ket{\varphi_{F,d}}=\hat U_F\ket{d}.
  \label{eq:rsu-frame-born}
\end{equation}
Thus, the prepared electronic state is unchanged; what changes is the determinant basis in which it is sampled.
In particular, configurations distributed over many low-probability canonical bitstrings can contribute
to a rotated determinant that is observed more readily in its own frame. The spread of a frame-dependent
sampling distribution can be characterized, for example, by its Shannon entropy,
\begin{equation} \eta_k(F) =-\sum_{d\in\mathcal{X}}P_{k,F}(d)\log P_{k,F}(d),
  \label{eq:rsu-entropy}
\end{equation}
or by the collision probability
\begin{equation} \pi_k(F)=\sum_{d\in\mathcal{X}}P_{k,F}(d)^2.
  \label{eq:rsu-collision}
\end{equation}
These quantities describe the frame-dependent distribution of measured
rotated determinants, not the distribution of canonical candidates after
unrotation. Orbital-frame choice provides a degree of freedom to change both
the probabilities and their concentration.
Frame quality is evaluated by how effectively each frame identifies useful canonical determinants and
improves energy convergence at a fixed total shot count, rather than by whether it increases or
decreases the sampling entropy.
\paragraph{Rotate--sample--unrotate selection.}
\indent
Let $\mathcal{F}$ denote the set of complementary orbital frames used for RSU, including the identity frame.
For each frame $F\in\mathcal{F}$, we prepare the reference state, apply the inverse orbital rotation,
and measure in the computational basis.
Let $n_{k,F,d}$ be the number of times the determinant $d$ is observed for state $k$ in frame $F$.
The rotated determinant associated with $d$ has the following expansion in the
canonical determinant basis,
\begin{equation}
  \ket{\varphi_{F,d}}
  =
  \sum_{x\in\mathcal{X}}
  A^{(F)}_{d,x}\ket{x}.
  \label{eq:rsu-canonical-expansion}
\end{equation}
The coefficients $A^{(F)}_{d,x}$ are obtained from determinants of the
corresponding submatrices of the orbital rotation matrix. When the orbital
rotation preserves spatial and spin symmetries, $A^{(F)}_{d,x}$ vanishes
unless $x$ and $d$ have the same spin-resolved electron occupations in each
irreducible representation.

We use the observations to assign each canonical determinant the RSU score
\begin{equation}
  \sigma_x
  =
  \sum_k
  \sum_{(F,d)\ {\rm observed}}
  n_{k,F,d}
  \left|A^{(F)}_{d,x}\right|^2.
  \label{eq:rsu-proposal-score}
\end{equation}
Direct evaluation of Eq.~\eqref{eq:rsu-proposal-score} for every
$x\in\mathcal{X}$ would require enumeration of the full canonical determinant
space. This enumeration can be avoided by generating a finite set of canonical
candidate determinants directly from the observed rotated determinants.

For this purpose, an observed record $(k,F,d)$ is selected with probability
$n_{k,F,d}/W$, where
\begin{equation}
  W=
  \sum_k
  \sum_{(F,d)\ {\rm observed}}
  n_{k,F,d}.
  \label{eq:rsu-total-count}
\end{equation}
A canonical determinant $x$ is then sampled with conditional probability
$\left|A^{(F)}_{d,x}\right|^2$. The resulting probability of generating $x$ is
\begin{equation}
  q_x
  =
  \frac{1}{W}
  \sum_k
  \sum_{(F,d)\ {\rm observed}}
  n_{k,F,d}
  \left|A^{(F)}_{d,x}\right|^2
  =
  \frac{\sigma_x}{W}.
  \label{eq:rsu-candidate-mixture}
\end{equation}

Canonical determinants with large RSU scores are preferentially generated
without enumerating the full determinant space or constructing a
statevector. After duplicate candidates are removed,
Eq.~\eqref{eq:rsu-proposal-score} is evaluated only for the resulting finite
candidate set $\mathcal C$. The candidate-generation procedure and its
computational details are described in Supporting Information,
Sec.~\ref{si:candidate-backtransform}.

As shown in Sec.~\ref{si:rsu-interference} of the Supporting Information,
Eq.~\eqref{eq:rsu-proposal-score} does not include interference between
distinct rotated determinants. Although these interference contributions are
generally nonzero, they are neglected when estimating the contribution of each
canonical determinant. In the Supporting Information, we also evaluate a
modified score that partially includes these contributions and find no
substantial change in energy convergence.

The RSU score $\sigma_x$ is used to select canonical determinants for the
subsequent selected-CI calculation. The selected determinant set is defined as
\begin{equation}
  \mathcal B_{\mathrm{RSU}}
  =
  \operatorname{Top}_{B}
  \{\sigma_x:x\in\mathcal C\},
  \label{eq:rsu-topb-impl}
\end{equation}
where $B$ is the maximum number of selected determinants. The score $\sigma_x$
is used only for determinant selection and does not represent a CI coefficient
or a Born probability.
For the identity frame, $A^{(I)}_{d,x}=\delta_{dx}$, so the construction reduces to conventional canonical QSCI.

\paragraph{Constructing orbital frames: token-cumulant natural orbitals.}
\label{par:tcno}
Natural orbitals, which diagonalize the one-particle reduced density matrix (1-RDM),
provide a natural choice of orbital frames by reflecting the electronic occupations of the state.
We consider three approaches to constructing such frames.
The first uses token-cumulant natural orbitals (TC-NO), which are obtained from the
known state-preparation circuit without measuring the 1-RDM.
Let $\boldsymbol{\tau}_k=(\tau_{k,1},\ldots,\tau_{k,L_k})$ specify an ordered sequence of single-
and double-excitation operators acting on the closed-shell reference $\ket{\Phi_k}$.
Although this sequence is obtained from GQE in the present calculations, the TC-NO construction
applies to any state-preparation circuit with known excitation operators, coefficients, and ordering.
Each excitation token specifies an excitation label $\mu_{k,\ell}$ and
a signed rotation angle $w_{k,\ell}$. The label identifies the spin
orbitals involved, while the angle specifies the magnitude and sign
of the corresponding rotation.

Let $\hat E_\mu$ be the corresponding excitation operator and
$\hat G_\mu=\hat E_\mu-\hat E_\mu^\dagger$ its anti-Hermitian generator. For
example, a single excitation uses
\begin{equation}
 \hat E_\mu=
 \hat a^\dagger_{a\sigma}\hat a_{i\sigma},
\end{equation}
whereas a double excitation uses
\begin{equation}
 \hat E_\mu=
 \hat a^\dagger_{a\sigma}
 \hat a^\dagger_{b\rho}
 \hat a_{j\rho}
 \hat a_{i\sigma}.
\end{equation}

The rotation angles are selected from a fixed discrete set
$\mathcal W$,
\begin{equation}
 w_{k,\ell}\in\mathcal W.
 \label{eq:tcno-token-weight}
\end{equation}
The angle magnitudes need not be identical for different tokens.
The angle set used in the present calculations is specified in
Sec.~\ref{sec:comp}. The circuit can be written as
\begin{equation}
 U(\boldsymbol{\tau}_k)=
 e^{w_{k,L_k}\hat G_{\mu_{k,L_k}}}
 \cdots
 e^{w_{k,1}\hat G_{\mu_{k,1}}}.
\end{equation}
Identity tokens contribute zero. To first order in the circuit weights,
\begin{equation}
 U(\boldsymbol{\tau}_k)\ket{\Phi_k}
 =
 \ket{\Phi_k}
 +
 \sum_{\ell}
 w_{k,\ell}
 \hat E_{\mu_{k,\ell}}\ket{\Phi_k}
 +
 O(w^2).
\end{equation}
The de-excitation operators annihilate the determinant reference.

For a determinant reference, repeated signed tokens are accumulated into
single- and double-excitation amplitudes. The occupied--virtual block retains
the linear single-excitation contribution, while the occupied--occupied and
virtual--virtual blocks include quadratic single- and double-amplitude
contractions. Their explicit definitions, opposite-spin index conventions,
same-spin antisymmetry, and factors correcting double counting are given in
Supporting Information, Sec.~\ref{si:tcno-construction},
Eqs.~\eqref{eq:si-tcno-singles}--\eqref{eq:si-tcno-oo}.
The resulting matrix is an approximate frame proposal, not the exact circuit
1-RDM; omitted mixed and gate-order corrections and the occupation limitations
are specified there. The circuit description is thus converted into an
approximate state-dependent one-particle matrix. Its spin-resolved form is diagonalized as
\begin{equation}
 \widetilde{\gamma}^{\sigma}_{k}
 R_{k,\sigma}
 =
 R_{k,\sigma}
 \Lambda_{k,\sigma},
 \label{eq:tcno-diagonalization}
\end{equation}
where the eigenvector matrices $R_{k,\sigma}$ define the TC-NO orbital frame
\begin{equation}
 F_k^{\mathrm{TC}}
 =
 (R_{k,\alpha},R_{k,\beta}).
\end{equation}
Only the known reference occupation and the ordered single- and
double-excitation circuit are used in this construction. Therefore, no
additional quantum measurement of the 1-RDM is required.

For a general seed state $\ket{\Psi_k^{\mathrm{seed}}}$, the leading correction
to the spin-resolved 1-RDM may instead be evaluated through the commutator
expression
\begin{equation}
 \widetilde{\gamma}^{\sigma}_{pq}
 =
 \gamma^{\sigma,\mathrm{seed}}_{pq}
 +
 \sum_{\ell}
 w_{k,\ell}
 \bra{\Psi_k^{\mathrm{seed}}}
 \left[
  \hat a^\dagger_{p\sigma}\hat a_{q\sigma},
  \hat G_{\mu_{k,\ell}}
 \right]
 \ket{\Psi_k^{\mathrm{seed}}}
 +
 O(w^2).
 \label{eq:seed-tcno-density}
\end{equation}
The general-seed form, including transition densities between seed
determinants, is detailed in Supporting Information,
Eq.~\eqref{eq:si-seed-tcno-density}; it is distinct from the
determinant-reference quadratic contractions. This form makes explicit that
the TC-NO construction is determined by the
reference state and the known excitation circuit, rather than by the method
used to optimize or generate that circuit.

\paragraph{State-targeted SCI-NO and RES-NO frames.}
A complementary route constructs orbital frames from a classically available
CI-like state. For
$\ket{\psi}=\sum_I c_I\ket{I}$, the spin-resolved 1-RDM is
\begin{equation}
 \gamma^\sigma_{pq}
 =
 \bra{\psi}
 \hat a^\dagger_{p\sigma}
 \hat a_{q\sigma}
 \ket{\psi},
 \qquad
 \gamma^\sigma R_\sigma
 =
 R_\sigma\Lambda_\sigma.
 \label{eq:orbital-frame-density}
\end{equation}
The eigenvectors $R_\sigma$ define the corresponding natural-orbital frame.
The preparation of the reference state and the construction of the orbital
frame are therefore separate steps.
TC-NO uses the known excitation circuit, whereas SCI-NO and RES-NO use information
obtained from an initial selected-CI calculation based on measurements 
in the canonical orbital basis using conventional QSCI.

For excited-state discovery, a ground-state-oriented or circuit-derived frame
need not be aligned with determinant directions that are missing from the
target excited state. We therefore introduce SCI-NO and RES-NO as
state-targeted alternatives to TC-NO.
Let $\mathcal B_0$ denote the initial set of canonical determinants obtained
from measurements in the canonical orbital basis, and let
\begin{equation}
 \ket{\psi_k^{(0)}}
 =
 \sum_{I\in\mathcal B_0}
 c_I^{(k)}\ket{I}
\end{equation}
be the normalized target state obtained by selected-CI diagonalization in
$\mathcal B_0$. For the ground state, the lowest singlet root in the target
spatial-symmetry sector is used without a fixed-ground orthogonality
constraint. For the target excited state, the lowest singlet root in the
target spatial-symmetry sector satisfying orthogonality to the fixed
approximate ground state is used.
SCI-NO is defined by diagonalizing the spin-resolved 1-RDM of
$\ket{\psi_k^{(0)}}$. The required matrix elements are evaluated classically
from the stored determinants and CI coefficients. Thus, SCI-NO does not
require a quantum measurement of the 1-RDM.
The CI wavefunction used to construct SCI-NO contains only determinants in $\mathcal B_0$,
and the 1-RDM to be diagonalized has the orbital dimension, not the dimension of the full determinant space.
This construction requires neither an FCI wavefunction nor an FCI-dimensional
Hamiltonian diagonalization.

RES-NO instead uses the residual component outside the initial selected-CI
space. In the present implementation, this component is restricted to the
target spatial symmetry and projected orthogonally to the fixed ground
state:
\begin{equation}
 \ket{r_k}
 =
 Q_{g,\mathrm{out}}
 Q_{\mathcal B_0}
 Q_k
 (\hat H-E_k^{(0)})
 \ket{\psi_k^{(0)}}.
 \label{eq:resno-residual}
\end{equation}
Here, $Q_k$ retains the target irreducible representation, and
$Q_{\mathcal B_0}$ removes components contained in the initial selected
determinant space $\mathcal B_0$. For the target excited state,
$Q_{g,\mathrm{out}}$ removes the direction associated with the fixed
ground state in the external space. For the ground-state calculation,
this fixed-ground projection is omitted. After normalization,
\begin{equation}
 \ket{\bar r_k}
 =
 \frac{\ket{r_k}}
 {\sqrt{\braket{r_k}{r_k}}}.
\end{equation}
The RES-NO frame is then obtained by diagonalizing the spin-resolved 1-RDM of
$\ket{\bar r_k}$.
Because the molecular Hamiltonian, the determinants in $\mathcal B_0$, and their CI coefficients
are already available from the initial selected-CI calculation, both the residual
and its 1-RDM can be constructed classically.
In particular, RES-NO requires neither quantum preparation of a residual state
nor diagonalization of the Hamiltonian in the external determinant space.
Only single and double substitutions connected to the determinants in $\mathcal B_0$
need to be generated to construct the unprojected residual; external determinants are not
recursively expanded.
The required fixed-ground projection may additionally
use the external support of the stored ground state. These determinants form a
classical auxiliary representation for the residual 1-RDM, not a variational
CI space to be diagonalized, and are not automatically added to the final
selected-CI support. Thus, constructing RES-NO is not an FCI solve and does
not add these external determinants directly to the final selected-CI space.

The set of external determinants connected to $\mathcal B_0$ can be larger
than the initial selected determinant space itself.
A memory-limited alternative screens candidates by Hamiltonian coupling to the initial
selected-CI coefficients, accumulates all pilot contributions for each
retained candidate, and then selects by residual weight. This limits retained
storage but approximates the exact residual and has not been used to validate
the reported curves. The coefficient formulas, candidate-selection procedure,
projection conditions, and classical costs are detailed in Supporting
Information, Sec.~\ref{si:classical-resources}.

The corresponding hardware workflow follows directly from this adaptive
construction.
The prepared reference state $\ket{\Psi_k}$ is first measured in the canonical
orbital basis using $N_{\mathrm{init}}$ shots. The resulting bitstrings define
an initial spin-completed set of canonical determinants $\mathcal B_0$.
The molecular Hamiltonian is constructed in this space, and selected-CI
diagonalization gives $\ket{\psi_k^{(0)}}$ using the state-specific
conditions defined above. The fixed-ground orthogonality constraint is
applied only to the target excited-state calculation.

SCI-NO is then obtained classically by diagonalizing the 1-RDM of $\ket{\psi_k^{(0)}}$.
For RES-NO, the known molecular Hamiltonian is applied to $\ket{\psi_k^{(0)}}$, the projectors
in Eq.~\eqref{eq:resno-residual} are imposed, and the 1-RDM of the normalized
residual is constructed classically.

The resulting orbital rotation is compiled into a one-body Gaussian circuit,
whose inverse is appended to the same state-preparation circuit. The remaining
$N_{\mathrm{prod}}$ shots are allocated to computational-basis measurements
in the identity and selected rotated frames. Finally, the frame-labeled
bitstrings are unrotated classically, the proposal score in
Eq.~\eqref{eq:rsu-proposal-score} is accumulated, and the final canonical
support is used for the orthogonal selected-CI energy evaluation.

Once the reference and pilot CI coefficients are available, SCI-NO and
RES-NO require neither Pauli-term measurements of $\hat H$ nor
quantum measurements of a 1-RDM. Their quantum overhead instead consists of a
second sampling stage, the additional orbital-frame circuits, the depth of
the corresponding one-body rotations, and
the division of a finite number of shots between the initial canonical-basis measurements
and the subsequent measurements in the selected orbital frames.
The present noiseless sampling study does not quantify hardware errors associated with these
orbital-rotation circuits.

\indent The algorithm is a hybrid quantum--classical selected-CI pipeline
(Figure~\ref{fig:pipeline-text}). Throughout, $\Hhat$ is the
second-quantized molecular Hamiltonian,
\begin{equation}
\Hhat=\sum_{pq}h_{pq}\,\crea{p}\anni{q}
      +\tfrac{1}{2}\sum_{pqrs}(pq\vert rs)\,\crea{p}\crea{r}\anni{s}\anni{q},
\label{eq:hamiltonian}
\end{equation}
with $p,q,r,s$ running over spin orbitals. We compute $K+1$ states labeled
$k=0,\dots,K$, with $k=0$ the ground state $S_{0}$;
$n_{\mathrm{state}}=K+1$.

\begin{figure}[t]
\centering
\setlength{\fboxsep}{6pt}

% Hanging-indented entry: wrapped lines begin at the same horizontal
% position as the first line after the specified indentation.
\newcommand{\FlowItem}[2]{%
  \par\noindent\hangindent=#1\hangafter=1\hspace*{#1}#2\par}

\fbox{\parbox{0.93\linewidth}{\small\ttfamily
\textbf{Stage 1: State preparation} $\to \ket{\Psi_k}$
(GQE or exact FCI control)\par
\FlowItem{2em}{[S$_0$] Ground-state GQE circuit}
\FlowItem{2em}{[S$_1$] Excited-state GQE circuit evaluated using the sampled
selected-CI energy}
\FlowItem{4em}{with the singlet and fixed-ground orthogonality conditions
imposed classically}
}}

\vspace{2pt}
$\downarrow$
\vspace{2pt}

\fbox{\parbox{0.93\linewidth}{\small\ttfamily
\textbf{Stage 2: Orbital-frame construction}\par
\FlowItem{2em}{Conventional QSCI: retain the canonical orbital basis}
\FlowItem{2em}{TC-NO case: construct the orbital frame from the known
state-preparation circuit}
\FlowItem{2em}{For SCI-NO and RES-NO cases: collect 10\% of the total shots in
the canonical orbital basis}
\FlowItem{4em}{$\to$ initial spin-completed support $\mathcal B_0$
$\to$ selected-CI state $\ket{\psi_k^{(0)}}$}
\FlowItem{2em}{SCI-NO: construct the orbital frame from the 1-RDM of
$\ket{\psi_k^{(0)}}$}
\FlowItem{2em}{RES-NO: construct the orbital frame from the 1-RDM of the
external residual}
}}

\vspace{2pt}
$\downarrow$
\vspace{2pt}

\fbox{\parbox{0.93\linewidth}{\small\ttfamily
\textbf{Stage 3: Rotate--sample--unrotate determinant discovery}\par
\FlowItem{2em}{Rotate: apply the inverse orbital rotation to
$\ket{\Psi_k}$}
\FlowItem{2em}{Sample: measure determinants in the rotated orbital frame}
\FlowItem{2em}{Unrotate: generate a finite set of candidates in the canonical
orbital basis using the orbital-transformation coefficients}
\FlowItem{2em}{Select: evaluate $\sigma_x$ only for the finite candidate set
(Eq.~\ref{eq:rsu-proposal-score})}
\FlowItem{4em}{$\to$ retain the highest-ranked candidates $+$ spin completion
$\to \mathcal B$}
\FlowItem{2em}{The full determinant space is not enumerated.}
}}

\vspace{2pt}
$\downarrow$
\vspace{2pt}

\fbox{\parbox{0.93\linewidth}{\small\ttfamily
\textbf{Stage 4: Orthogonal selected-CI diagonalization in $\mathcal B$}\par
\FlowItem{2em}{[S$_0$] Lowest singlet root
(Eq.~\ref{eq:mode4A})}
\FlowItem{2em}{[S$_1$] Lowest singlet root orthogonal to the fixed
$\ket{g}$ (Eq.~\ref{eq:fixed-ground-readout})}
}}

\caption{QSCI/RSU workflow for the ground state (S$_0$) and the target
excited singlet (S$_1$). The four framed stages show state preparation,
orbital-frame construction, finite-candidate RSU determinant discovery,
and final selected-CI diagonalization. For SCI-NO and RES-NO only, 10\% of
the total shots are first collected in the canonical orbital basis; these
shots are included in the total shot count. TC-NO is constructed from the
known state-preparation circuit. Conventional QSCI follows the canonical-
basis path without orbital rotation or RSU candidate generation. For S$_1$,
$\ket{g}$ is a fixed selected-CI ground-state approximation that is not
reoptimized.}
\label{fig:pipeline-text}
\end{figure}

\subsection{GQE-based excited-state reference  preparation}
\label{sec:reference-preparation}
We extend GQE from ground-state preparation to the preparation of sampling
states for excited-state QSCI. The purpose is to generate a quantum state whose
computational-basis measurements contain determinants that form an accurate
selected-CI space for the target excited state.

For a target state $k$, GQE generates an ordered sequence of excitation tokens
\begin{equation}
 \boldsymbol{\tau}_k=(\tau_{k,1},\ldots,\tau_{k,L_k}),
\end{equation}
which defines the state-preparation circuit
\begin{equation}
 \ket{\Psi_k(\boldsymbol{\tau}_k)}
 =U(\boldsymbol{\tau}_k)\ket{\Phi_k},
 \qquad
 U(\boldsymbol{\tau}_k)
 =U_{\tau_{k,L_k}}\cdots U_{\tau_{k,1}}.
 \label{eq:multistate-gqe-reference}
\end{equation}
The token set is chosen so that particle number, spin, and the target spatial
symmetry are preserved. Ground- and excited-state circuits are generated
separately because the determinant distributions required for their subsequent
selected-CI calculations are different.

For the ground state, the GQE circuit is prepared from the closed-shell
reference determinant and is optimized with the ground-state objective used in
the preceding GQE calculation.
For the target excited singlet considered in this work, the starting state
$\ket{\Phi_1}$ is a spin-adapted configuration state function constructed from
a single excitation in the selected spatial-symmetry sector and its
spin-flipped counterpart. The relative phase is chosen to form a singlet.
When the symmetry of the target excited state is not known in advance,
separate calculations can be performed for the relevant symmetry sectors.
The excited-state GQE circuit is then optimized using the selected-CI energy
obtained from determinants sampled from each candidate token sequence.

\paragraph{Sampled selected-CI objective for excited-state GQE.}
For a candidate sequence $\boldsymbol{\tau}$, the quantum circuit prepares
the candidate state $\ket{\Psi_1(\boldsymbol{\tau})}$ for the target excited
singlet and measures it in the computational basis.
Let $n_I(\boldsymbol{\tau})$ be the number of observations of determinant
$\ket{I}$. Only determinants with the required electron number, spin
projection, and spatial symmetry are retained. They are ordered according to
$n_I(\boldsymbol{\tau})$, and the leading determinants define the observed set
\begin{equation}
  \mathcal D(\boldsymbol{\tau})
  =
  \operatorname{Top}_{M}
  \left\{
  \ket{I}
  \mid
  n_I(\boldsymbol{\tau})>0
  \right\}.
  \label{eq:gqe-sampled-determinants}
\end{equation}
Here, $\operatorname{Top}_{M}$ retains at most $M$ distinct observed
determinants with the largest observation counts before spin completion.
The selected set is then completed under spin exchange to form the spin-complete
space $\mathcal S(\boldsymbol{\tau})$, whose dimension may exceed $M$.
In the present calculations, we set
$M=50$, which limits the number of observed determinants retained before spin
completion. This value is distinct from the maximum of 2000 canonical
determinants adopted for the final QSCI calculations.

The Hamiltonian and spin matrices are constructed classically only within
$\mathcal S(\boldsymbol{\tau})$
\begin{equation}
 H_{IJ}(\boldsymbol{\tau})
 =\bra{I}\hat H\ket{J},
 \qquad
 S_{IJ}(\boldsymbol{\tau})
 =\bra{I}\hat S^2\ket{J},
 \qquad
 \ket{I},\ket{J}\in\mathcal S(\boldsymbol{\tau}).
 \label{eq:gqe-sampled-support-matrices}
\end{equation}
Within the singlet subspace, the lowest root satisfying the fixed-ground
condition defines
\begin{equation}
 \mathcal L_1(\boldsymbol{\tau})
 =
 \min_{\boldsymbol c\ne 0}
 \frac{\boldsymbol c^\dagger
 H(\boldsymbol{\tau})\boldsymbol c}
 {\boldsymbol c^\dagger\boldsymbol c},
 \label{eq:gqe-sampled-ci-loss}
\end{equation}
where the minimization is restricted by
\begin{equation}
  \boldsymbol c^\dagger S(\boldsymbol{\tau})\boldsymbol c = 0,
  \quad
  \braket{g}{\psi(\boldsymbol{\tau})} = 0,
  \quad
  \ket{\psi(\boldsymbol{\tau})}
  = \sum_{I\in\mathcal S(\boldsymbol{\tau})} c_I\ket{I}.
\label{eq:gqe-sampled-ci-constraints}
\end{equation}

Here $\ket{g}$ is the fixed approximate ground state. In the present
calculations, the target excited states and the fixed ground state belong to
different spatial irreducible representations. Their overlap therefore
vanishes by symmetry, although the fixed-ground condition is retained in the
selected-CI calculation.

The GQE policy assigns the reward
\begin{equation}
 R(\boldsymbol{\tau})=-\mathcal L_1(\boldsymbol{\tau})
 \label{eq:gqe-sampled-ci-reward}
\end{equation}
and updates the probability of generating token sequences.
Each candidate GQE circuit is evaluated using only the determinants sampled
from that circuit. After learning, the circuit with the lowest selected-CI
energy is used to prepare the excited-state sampling state for the subsequent
QSCI and RSU calculations.

This construction separates the quantum and classical tasks. The quantum
calculation prepares the GQE state and measures determinants in an orbital
basis. It does not require measurements of the Hamiltonian, $\hat S^2$, or the
overlap with the fixed ground state. The classical calculation ranks the
observed determinants, completes the selected space under spin exchange, and
solves the selected-CI eigenvalue problem within that finite space. The
resulting GQE state supplies the determinant samples used for QSCI in the
canonical orbital basis and for RSU in the orbital frames introduced in
Sec.~\ref{sec:stage3-discovery}.

\subsection{Canonical candidate selection}
\label{sec:candidate-refinement}
\paragraph{Canonical bitstring candidates.}
Conventional QSCI selects canonical determinants according to their
measurement counts. In RSU, observations from all retained orbital frames are
instead unrotated and accumulated into the canonical proposal score $\sigma_x$
defined in Eq.~\eqref{eq:rsu-proposal-score}.
Candidates are considered in descending order of their scores. Each
candidate is accepted together with all additional determinants required
for spin completion only if the resulting selected-CI dimension does not
exceed $B$. The final space is therefore spin-complete and contains at
most $B$ canonical determinants, without subsequent truncation of
individual determinants.
Both methods therefore produce a selected space in the same orthonormal
canonical determinant basis, but differ in how its candidates are identified.

\paragraph{State-targeted selection.}
The RSU procedure defined above ranks canonical candidates using
$\sigma_x$. As an optional extension, the residual of an initial
selected-CI state can be used to rerank the same finite candidate set
according to its relevance to a target electronic state. This extension
changes candidate ranking, not the orbital-frame construction or the
candidate-generation procedure.

Let $\mathcal C$ denote this finite candidate set. We first construct an
initial canonical determinant space $\mathcal B_0$ from the highest-ranked
identity-frame observations. Let $\ket{\psi_k^{(0)}}$ and $E_k^{(0)}$ denote
the Ritz state and its energy obtained in $\mathcal B_0$. For each candidate
$x\in\mathcal C\setminus\mathcal B_0$, we define
\begin{equation}
\begin{aligned}
  r_{k,x}
  &=
  \bra{x}
  \left(\hat H-E_k^{(0)}\right)
  \ket{\psi_k^{(0)}},\\
  g_{k,x}
  &=
  \frac{|r_{k,x}|^2}
  {|H_{xx}-E_k^{(0)}|+\epsilon},\\
  H_{xx}
  &=
  \bra{x}\hat H\ket{x}.
\end{aligned}
\label{eq:candidate-residual}
\end{equation}
Here, $\epsilon>0$ regularizes a small denominator. The residual-targeted
score is then defined for the same finite candidate set as
\begin{equation}
  t_{k,x}
  =
  \sigma_x
  (g_{k,x}+f)^\alpha,
  \quad
  x\in\mathcal C\setminus\mathcal B_0,
  \label{eq:candidate-residual-score}
\end{equation}
where $f>0$ is a score floor and $\alpha>0$ controls the influence of the
residual.

For state-averaged selection, $g_{k,x}$ is replaced by
$\sum_k\omega_k g_{k,x}$, where $\omega_k$ are nonnegative state weights.

This procedure only reranks the determinants in the finite candidate set
$\mathcal C$ and does not search the full determinant space $\mathcal X$.
We refer to selection using $\sigma_x$ alone as frequency-ranked RSU and to
selection using Eq.~\eqref{eq:candidate-residual-score} as residual-targeted
RSU (RT-RSU).
In both QSCI+res and RT-RSU, the initial space $\mathcal B_0$ contains
$|\mathcal B|/2$ determinants selected from identity-frame observations. The
remaining determinants are selected using the residual-targeted score in
Eq.~\eqref{eq:candidate-residual-score}. As a control, we apply this score to
the finite candidate set obtained only from identity-frame measurements. We
refer to this procedure as QSCI+res. In RT-RSU, the same score is instead
applied to the finite candidate set generated using all retained orbital
frames.

\subsection{Final orthogonal selected-CI diagonalization}
\label{sec:stage4}
Let $\mathcal B\subseteq\mathcal X$ denote the canonical determinant space
selected by QSCI or RSU. The Hamiltonian and squared-spin matrices in this
space are
\begin{equation}
  H_{xy}^{\mathcal B}
  =
  \bra{x}\hat H\ket{y},
  \qquad
  Q_{xy}^{\mathcal B}
  =
  \bra{x}\hat S^2\ket{y},
  \qquad
  x,y\in\mathcal B.
  \label{eq:final-ci-matrices}
\end{equation}

Because the canonical determinants are orthonormal,
$\braket{x}{y}=\delta_{xy}$, no overlap matrix or generalized eigenvalue
problem is required for the selected-CI calculation.
The same canonical-basis formulation is used for ground- and excited-state
calculations, although the selected determinant space may differ between
target states. To identify singlet roots in a selected space that may also
contain other spin sectors, we diagonalize the optionally spin-shifted
matrix
\begin{equation}
  \sum_{y\in\mathcal{B}} (H_{xy}^{\mathcal{B}}+\mu Q_{xy}^{\mathcal{B}})c_{y\nu} = \lambda_\nu c_{x\nu}, \qquad \ket{\Phi_\nu^{\mathcal{B}}} = \sum_{x\in\mathcal{B}}c_{x\nu}\ket{x},
  \label{eq:mode4A}
\end{equation}
where $\mu\geq 0$ is the spin-shift parameter. For each normalized eigenvector, we
evaluate
\begin{equation} s_\nu = \bra{\Phi_\nu^{\mathcal{B}}}\hat S^2 \ket{\Phi_\nu^{\mathcal{B}}}, \qquad E_\nu = \bra{\Phi_\nu^{\mathcal{B}}}\hat H \ket{\Phi_\nu^{\mathcal{B}}}.
  \label{eq:final-ci-physical-energy}
\end{equation}
The reported energy is $E_\nu$, not the shifted eigenvalue $\lambda_\nu$. Roots satisfying
$|s_\nu|\leq s_{\mathrm{tol}}$ are identified as singlet-like, where $s_{\mathrm{tol}}$ is the specified
spin tolerance.
When a common selected space is used for a conventional two-root analysis,
the two lowest qualifying singlet roots are assigned to S$_0$ and S$_1$.
For the target excited-state calculations reported here, the energy is
instead evaluated under the fixed-ground orthogonality constraint
described below.

\paragraph{Excited-state energy evaluation with an orthogonality constraint.}
\label{par:excited-orthogonal-readout}
For a selected determinant space $\mathcal B\subseteq\mathcal X$, let
$D_{\mathcal B}$ embed this space into the complete canonical determinant
space $\mathcal X$. We define
\begin{equation}
  H_{\mathcal B}
  =
  D_{\mathcal B}^\dagger H D_{\mathcal B},
  \qquad
  T_{\mathcal B}
  =
  D_{\mathcal B}^\dagger S^2 D_{\mathcal B},
  \qquad
  b_{\mathcal B}
  =
  D_{\mathcal B}^\dagger g.
\end{equation}
Here, $H_{\mathcal B}$ is the Hamiltonian in the selected space,
$T_{\mathcal B}$ is used to identify its singlet subspace, and
$b_{\mathcal B}$ represents the fixed ground state in the same space.
Let $V_{\mathcal B}$ span the singlet subspace identified from
$T_{\mathcal B}$. To impose orthogonality to the fixed ground state, let
$Z_{\mathcal B}$ span the nullspace of
$b_{\mathcal B}^\dagger V_{\mathcal B}$.

We then solve
\begin{equation}
 Z_{\mathcal B}^\dagger V_{\mathcal B}^\dagger H_{\mathcal B} V_{\mathcal B} Z_{\mathcal B} v=E_\perp v,\qquad |\psi_{\mathcal B}\rangle=D_{\mathcal B} V_{\mathcal B} Z_{\mathcal B} v.
\label{eq:fixed-ground-readout}
\end{equation}
The lowest reduced root is used.
The ground state $\ket{g}$ is kept fixed and used only to impose the
orthogonality constraint.
If $\ket{g}$ is the exact ground state, minimizing the energy under the
orthogonality constraint gives a variational upper bound to the first excited
state energy. When $\ket{g}$ is approximate, orthogonality to $\ket{g}$ does
not completely remove the exact ground-state component. The resulting energy
is therefore not guaranteed to be an upper bound to the exact first excited
state energy.

\section{Computational Details}
\label{sec:comp}
We first describe the computational implementation used in this work.
Molecular one- and two-electron integrals were obtained using PySCF~\cite{sun2018pyscf}.
The Hamiltonian used in the QSCI calculations was constructed from these integrals, and the fermionic
operators were mapped to qubit operators using the Jordan--Wigner transformation implemented in
OpenFermion~\cite{OpenFermion}. All calculations employed the STO-3G basis and an active space
selected using automated construction of molecular valence active spaces (AVAS)~\cite{AVAS2017}.
Noiseless statevector simulations were performed using Qulacs~\cite{Qulacs2021}.

For GQE-based state preparation, the circuit generator was implemented as
a decoder-only causal Transformer with two layers. The ground-state model
used two attention heads and an embedding dimension of 32, whereas the
excited-state model used four attention heads and an embedding dimension
of 64.
The model used learned
token and positional embeddings. Each Transformer block consisted of
pre-layer normalization, multi-head causal self-attention, and a feed-forward
network with a GELU activation and an intermediate dimension four times the
embedding dimension. Residual connections were applied around the attention
and feed-forward modules. The model generated excitation-operator sequences
autoregressively, with each operator sampled from the conditional distribution
defined by the preceding tokens.
The Transformer parameters were optimized by group-relative policy optimization
(GRPO)~\cite{GRPO}, using objective values evaluated by noiseless statevector
simulations in Qulacs. 

The ground-state model used six discrete rotation angles for each of the
119 excitation generators. For the excited-state model, the rotation
angles were selected from 14 values: $\pm0.005$, $\pm0.01$, $\pm0.02$,
$\pm0.05$, $\pm0.1$, $\pm0.2$, and $\pm0.4$ rad.
We use two operator pools: one contains 29 particle--hole single- and
double-excitation generators relative to the Hartree--Fock determinant (HF pool),
and the other contains 119 symmetry-allowed, spin-adapted generators that also include orbital transitions not restricted to
occupied--virtual pairs (generalized pool). 

For N$_2$, the test molecule in this study, we use a 16-qubit $(10e,8o)$ active
space selected using AVAS~\cite{AVAS2017}, with only the N $1s$ core orbitals frozen.
For the ground-state calculations, three independently trained GQE
circuits were used at each geometry. Each circuit started from the
Hartree--Fock determinant and applied 64 excitation-operator tokens.
The ground-state vocabulary contained 715 tokens: one beginning-of-sequence
token and 119 excitation generators combined with six discrete rotation
angles.
For the excited-state calculations, the circuits selected using independent
validation selected-CI energies contained 16 and 64 tokens at
$R=1.1$ and $1.8$~\AA, respectively.

For the ground-state calculations, three independent GQE training runs were
performed at each geometry, each for 200 epochs.
For the excited-state calculations, curriculum training increased the
circuit length through 4, 8, 16, 32, and 64 excitation-operator tokens,
with 60 training rounds at each length. Each round used 32 candidate
circuits, each evaluated with 512 measurement shots in the canonical
orbital basis. At most 50 distinct determinants were retained according
to their observation counts before spin completion; the completed space
could therefore exceed 50 determinants. The resulting constrained
selected-CI energy was used to train the Transformer, with four GRPO
updates per round. The final circuit was selected using independently
evaluated validation selected-CI energies and kept fixed for the
subsequent QSCI/RSU calculations.

The space with $N_\alpha=N_\beta=5$ contains $\binom{8}{5}^{2}=3136$ determinants.
The GQE model was trained independently at each molecular geometry.
The determinant set was supplemented with spin-adapted, closed-shell paired-double seed determinants,
defined as follows:
\begin{equation}
\mathcal{D}_{\mathrm{PD}}=\Bigl\{\,\bigl(\occ_{0}\!\setminus\!\{o_1,\dots\}\cup\{a_1,\dots\}\bigr)^{\otimes\{\alpha,\beta\}}
\;:\;\{(o_r\!\to\!a_r)\}_{r=1}^{m},\ m\le m_{\max}\,\Bigr\},
\label{eq:paired-doubles}
\end{equation}
each a closed-shell singlet by construction.
For the main-text $S_1$ comparisons, both the FCI-prepared and GQE-prepared states
are sampled with total shot counts from 200 to 20,000. Additional FCI-prepared
$S_1$ measurements up to 100,000 shots are used only for the amplitude-based
scoring analysis in the Supporting Information. The GQE-prepared states are
kept unchanged throughout these sampling comparisons. For SCI-NO and RES-NO, 10\% of
the total shots are used for initial measurements in the canonical orbital basis to
construct the orbital frames, and each calculation is repeated with eight measurement random seeds.
For the GQE-prepared calculations, FCI results were used only to evaluate the accuracy.

The present calculations are small-active-space, noiseless statevector
simulations intended as a proof of concept for RSU. Validation on larger
active spaces and noisy quantum hardware is left for future work.

% ======================================================================
\section{Results}
\label{sec:results}
% ======================================================================
\subsection{Shot-efficient determinant discovery by RSU for the ground state}
\label{sec:res-rsu-n2}

To remove the effect of state-preparation error and assess the sampling efficiency of RSU,
we first use the exact active-space FCI ground-state wavefunction as the sampling input.
We compare conventional QSCI (identity-frame QSCI) with RSU using SCI-NO and RES-NO frame rotations,
and additionally with TC-NO for the GQE-prepared states.
These calculations use the same prepared wavefunction, total number of shots, canonical Hamiltonian,
spin-symmetry treatment, and selected-CI diagonalization procedure.
They differ only in the orbital frames used for measurement and the corresponding
transformation of the observed determinants to the canonical basis.
In practical calculations, wavefunctions prepared by methods such as VQE or GQE generally
differ from the exact eigenstate. We therefore follow the FCI-prepared calculations
with GQE-prepared calculations as a representative test of RSU with approximate wavefunctions.
The same design is used for the excited state in Sec.~\ref{sec:res-rsu-s1}.

\paragraph{Determinant discovery from the exact FCI state.}

The sampling input is the exact active-space FCI ground-state wavefunction,
$S_0(A_g)$, of N$_2$/STO-3G in the $(10e,8o)$ active space at bond lengths $R=1.1$ and $1.8$~\AA.
Conventional QSCI measures only in the canonical orbital basis, without RSU. TC-NO is not applied to
the FCI-prepared state because its orbital frame is constructed from the ordered excitation operators
and rotation angles of a known state-preparation circuit, which are not available for the exact FCI wavefunction.
For SCI-NO and RES-NO, an initial selected-CI calculation is performed using measurements
in the canonical basis. This initial step uses $\lfloor N/10\rfloor$ shots, included in the total
shot count $N$, and at most 170 determinants. Both frames are constructed from the same
initial calculation.
SCI-NO is obtained by diagonalizing the 1-RDM of the ground-state
selected-CI wavefunction obtained from the initial measurements. RES-NO
instead uses the residual component outside the initial selected-CI space.
This residual is obtained by applying $\hat H-E_0^{(0)}$ to the selected-CI
wavefunction and normalizing the resulting
state. The RES-NO frame is defined by diagonalizing the 1-RDM of this
normalized residual.

\begin{figure}[t]
  \centering
  \includegraphics[width=\linewidth]{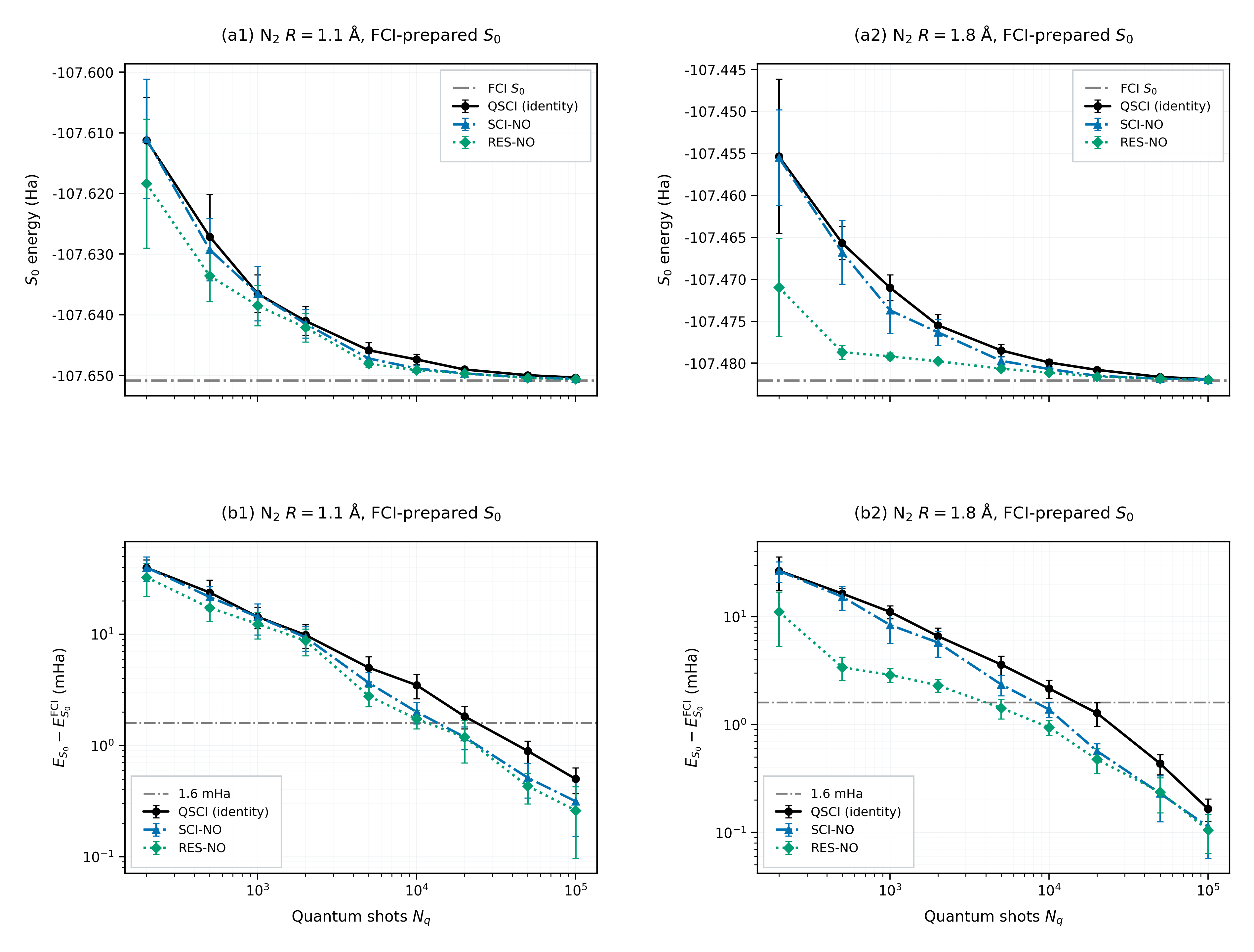}
   \caption{Finite-shot convergence of the FCI-prepared ground state of N$_2$/STO-3G $(10e,8o)$
    using candidate-restricted RSU. The target cap, defined as the maximum number
    of canonical determinants retained in the selected-CI space, is set to 2000.
    Upper panels: selected-CI energy; dash-dotted lines mark the active-space FCI energies.
    Lower panels: absolute error relative to FCI on a logarithmic scale; the dash-dotted line
    marks $1.6\mHa$. Left: $R=1.1$~\AA; right: $R=1.8$~\AA. QSCI uses the identity frame;
    SCI-NO and RES-NO use the state-targeted frames defined in Sec.~\ref{par:tcno}.
    For SCI-NO and RES-NO, the initial canonical-basis shots are included in the total number of shots.
    Symbols are means over eight measurement seeds, and error bars are sample standard deviations
    after averaging the classical repetitions within each measurement seed.}
  \label{fig:rsu-matched-n2}
\end{figure}

The remaining shots are distributed equally among the measurement frames.
Canonical candidates are generated from the observed rotated determinants and
ranked using the RSU score in Eq.~\eqref{eq:rsu-proposal-score}. The score is
evaluated only for this finite candidate set, which is then completed under
spin exchange.
QSCI, SCI-NO, and RES-NO use the same Hamiltonian and determine the lowest $A_g$ singlet
by selected-CI diagonalization.
The target cap specifies the maximum number of canonical determinants retained
in the selected-CI space and is set to 2000. Figure~\ref{fig:rsu-matched-n2}
includes nine total shot counts from $200$ to $10^{5}$.
Each point is the mean over eight measurement seeds,
and the error bars are sample standard deviations.

QSCI, SCI-NO, and RES-NO approach the FCI energy as the number of shots increases, but the number of measurements required to reach a given accuracy differs substantially (Fig.~\ref{fig:rsu-matched-n2}).
At $R=1.8$~\AA, where strong static correlation distributes
the ground-state wavefunction over many determinants, the benefit of RES-NO is particularly
clear at small shot counts. With 500 shots, the mean error decreases from $16.38\mHa$ for QSCI
to $3.39\mHa$ for RES-NO.
RES-NO first reaches a mean error below $1.6\mHa$ (chemical accuracy) at 5000 shots,
SCI-NO at 10,000 shots,
and QSCI at 20,000 shots. At $10^{5}$ shots, the mean errors are $0.165\mHa$ for QSCI, $0.113\mHa$
for SCI-NO, and $0.105\mHa$ for RES-NO.

At $R=1.1$~\AA, SCI-NO and RES-NO likewise reach the same accuracy with fewer measurements.
Both first reach a mean error below $1.6\mHa$ at 20,000 shots, whereas QSCI requires 50,000 shots.
At $10^{5}$ shots, the mean errors are $0.261\mHa$ for RES-NO and $0.315\mHa$ for SCI-NO, compared
with $0.500\mHa$ for QSCI.

For the two bond lengths and nine shot counts tested, RES-NO gives a lower mean energy error
than QSCI in all 18 conditions and than SCI-NO in 17. These results show that the SCI-NO and RES-NO frames,
constructed from the initial selected-CI wavefunction and its external residual, provide more
effective measurement bases than the canonical orbital basis.

\paragraph{Determinant discovery from the GQE-prepared state.}
We next repeat the comparison using ground-state wavefunctions prepared by independently
trained GQE calculations.
The sampling wavefunction is replaced, while the Hamiltonian and measurement conditions remain unchanged.
These include an upper limit of 2000 selected canonical determinants, nine total shot counts from 200 to 100,000,
and eight measurement seeds shared by all frame rotation methods. For SCI-NO and RES-NO, the initial canonical-basis
measurements used to construct the orbital frames account for 10\% of the total shots.
Spin-symmetry completion and the determination of the lowest $A_g$
singlet by selected-CI diagonalization are also unchanged.
For each measurement seed, the energies obtained from the three independently
trained GQE circuits are first averaged. The reported mean and sample standard
deviation are then calculated from the resulting eight measurement-seed
averages.

For the RSU calculations, the finite candidate-generation procedure of
Sec.~\ref{si:candidate-exact-closure} is used. The RSU score is evaluated only
for the generated candidate determinants and not for every determinant in the
fixed-electron-number space.

\begin{figure}[t]
\centering
\includegraphics[width=\linewidth]{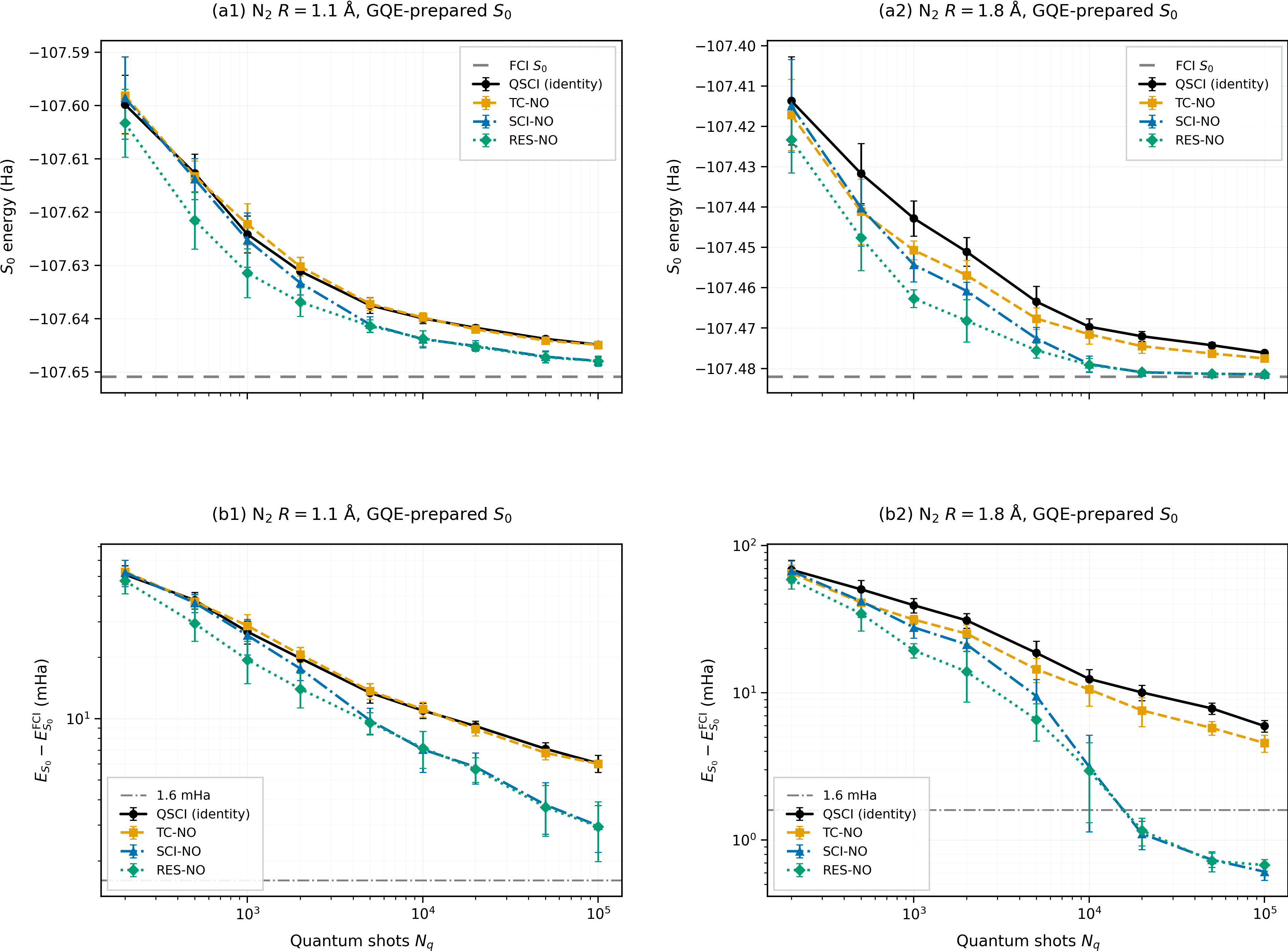}
\caption{Finite-shot convergence of the generalized-pool GQE-prepared
$S_0$ states of N$_2$/STO-3G $(10e,8o)$ using candidate-restricted RSU. The
target cap, defined as the maximum number of canonical determinants retained
in the selected-CI space, is set to 2000. Upper panels show the selected-CI
$S_0$ energy, and dash-dotted lines mark the active-space FCI energies. Lower
panels show the absolute error relative to FCI on a logarithmic scale, and the
dash-dotted line marks $1.6\mHa$. Left panels show $R=1.1$~\AA, and right
panels show $R=1.8$~\AA. QSCI uses the identity frame, whereas TC-NO, SCI-NO,
and RES-NO use the orbital frames defined in Sec.~\ref{par:tcno}.
The total number of shots includes the initial canonical-basis measurements used to
construct the SCI-NO and RES-NO frames.
Symbols are means over eight measurement seeds after averaging
the three independently trained GQE references within each measurement seed,
and error bars are sample standard deviations.}
\label{fig:rsu-gqe-s0}
\end{figure}

Figure~\ref{fig:rsu-gqe-s0} shows that the state-targeted frames retain their
finite-shot advantage for GQE wavefunction inputs when the RSU score is evaluated
on a finite candidate set. At $R=1.1$~\AA, the mean errors at $10^5$ shots are
$6.03$, $5.99$, $2.97$, and $2.94\mHa$ for QSCI, TC-NO, SCI-NO, and RES-NO,
respectively. At this geometry, SCI-NO and RES-NO improve on QSCI, although none
of the methods reaches chemical accuracy within the tested shot range.

At $R=1.8$~\AA, the mean errors at $10^5$ shots are $5.94\mHa$ for QSCI,
$4.56\mHa$ for TC-NO, $0.61\mHa$ for SCI-NO, and $0.68\mHa$ for RES-NO.
SCI-NO and RES-NO first reach a mean error below $1.6\mHa$ at 20,000 shots.
These results show that the determinant-discovery advantage of the
state-targeted rotated frames persists when the exact FCI sampling state is
replaced by an approximate GQE-prepared state. The larger absolute errors
relative to the FCI-prepared results reflect the remaining state-preparation
error in the GQE reference.

\paragraph{Origin of the enhanced determinant discovery by RSU.}
The FCI-prepared control isolates the measurement-frame effect from
state-preparation error and therefore provides the clearest setting in which
to identify the origin of the improvement.

The improved energy convergence is accompanied by more effective discovery of canonical
determinants entering the final selected-CI calculation. 

This improved determinant discovery arises because a determinant measured in
a rotated orbital basis is expanded over multiple determinants in the
canonical orbital basis. A rotated-frame measurement can therefore generate
canonical candidates that are difficult to observe directly in conventional
QSCI. RSU can thus provide a broader set of candidates for the selected-CI
calculation at the same number of measurements.

The principal advantage of RSU is the discovery of a broader set of important
canonical determinants from a finite number of measurements. The weights in
Eq.~\eqref{eq:rsu-proposal-score} are effective for identifying additional
candidate determinants, but they do not directly measure the energetic
importance of each determinant in a fixed-size selected-CI space. Improved
candidate discovery therefore does not necessarily imply a better ranking when
the number of retained determinants is fixed.

The GQE-prepared results support the same interpretation. Replacing the exact
FCI sampling input with an approximate GQE state changes the available
determinant distribution and increases the remaining state-preparation error,
but it does not change the role of RSU. In both cases, the improvement arises
from recovering useful canonical determinants through measurements in
complementary orbital frames rather than from a different final CI treatment.

\subsection{Shot-efficient determinant discovery by RSU for the excited state}
\label{sec:res-rsu-s1}
As in the ground-state calculations, we first use the exact FCI wavefunction
of the target $S_1$ state to remove state-preparation error and isolate the
effect of the orbital frame. We then repeat the comparison using a
GQE-prepared state.

For both the FCI- and GQE-prepared states, we compare conventional QSCI with SCI-NO and
RES-NO RSU. For the GQE-prepared state, TC-NO RSU is also included, with its frame constructed
from the GQE circuit information.
The RSU calculations evaluate only the
explicitly generated finite candidate sets and do not enumerate the full
determinant space.

\paragraph{Determinant discovery from the exact FCI state.}
We consider N$_2$/STO-3G in the $(10e,8o)$ active space at $R=1.1$ and
$1.8$~\AA. The targets are the lowest $B_{2g}$ singlet at $R=1.1$~\AA\ and
the lowest $A_u$ singlet at $R=1.8$~\AA, denoted as the $S_1$ states in the
present symmetry branches.
The exact FCI states are used only to generate determinant samples.
Conventional QSCI measures in the canonical orbital basis, whereas SCI-NO
and RES-NO use the state-targeted orbital frames defined in
Sec.~\ref{par:tcno}.
As in the ground-state calculations, TC-NO is not applied because the exact FCI state
is not prepared by a known excitation circuit.
The total number of shots are 200, 500, 1000, 2000, 5000,
and 20,000.
For SCI-NO and RES-NO, the first 10\% of the total shots are used for
conventional QSCI measurements in the canonical orbital basis to construct the
orbital frame. These shots are included in the total number of shots, $N$.

For RSU, canonical candidates are generated from the observed rotated
determinants, and the score in Eq.~\eqref{eq:rsu-proposal-score} is evaluated
only for this finite candidate set. The classical calculation therefore does
not scan the full fixed-electron-number determinant space. The target cap,
defined as the maximum number of canonical determinants retained in the
selected-CI space, is set to 2000. Each point is obtained from eight
measurement seeds.

For each method, orbital frame, molecular geometry, and total number of shots,
determinant sampling was performed independently eight times.
For each of these eight determinant-sampling runs, the stochastic RSU
candidate-selection procedure was repeated three times with different
random-number sequences, and the resulting three energies were averaged.
The reported mean and sample standard deviation were then calculated from
the eight independently sampled results.

\begin{figure}[t]
\centering
\includegraphics[width=\linewidth]{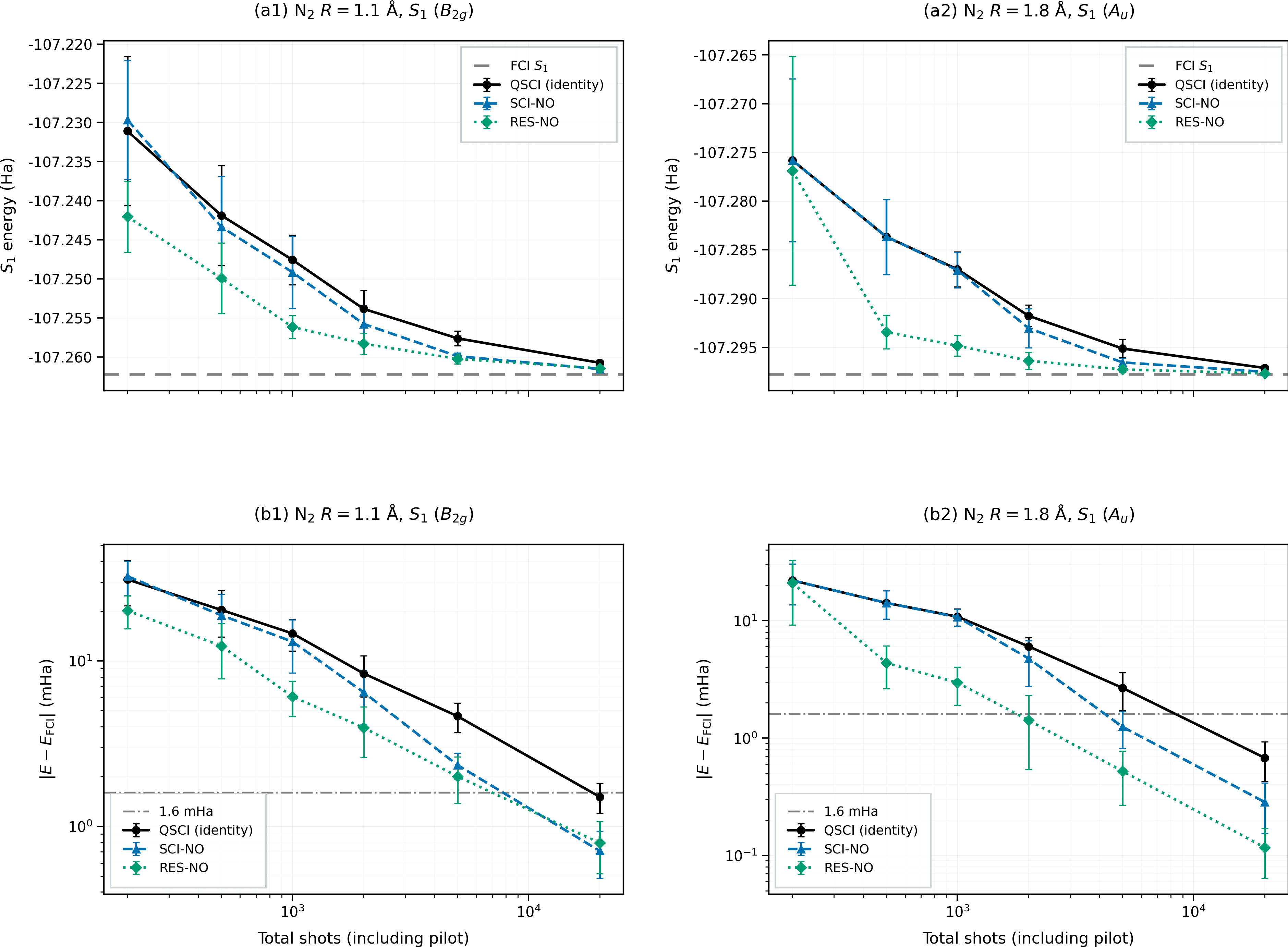}
\caption{Finite-shot convergence of the FCI-prepared $S_1$ states of
N$_2$/STO-3G $(10e,8o)$ using candidate-restricted RSU. The target cap,
defined as the maximum number of canonical determinants retained in the
selected-CI space, is set to 2000. Upper panels show the selected-CI $S_1$
energy, and dashed lines mark the corresponding active-space FCI energies.
Lower panels show the absolute error relative to FCI on a logarithmic scale,
and the dash-dotted line marks $1.6\mHa$. Left panels show the lowest
$B_{2g}$ singlet at $R=1.1$~\AA, and right panels show the lowest $A_u$
singlet at $R=1.8$~\AA. QSCI uses the identity frame, whereas SCI-NO and
RES-NO use the state-targeted orbital frames defined in Sec.~\ref{par:tcno}.
The total number of shots includes the initial canonical-basis measurements used to
construct the SCI-NO and RES-NO frames.
Symbols are means over eight measurement seeds, and
error bars are sample standard deviations. For RSU, the three classical-seed
results are averaged within each measurement seed before the reported
statistics are calculated.}
\label{fig:rsu-s1-frame-comparison}
\end{figure}

Figure~\ref{fig:rsu-s1-frame-comparison} shows that the selected-CI energies
approach the FCI reference as the number of measurements increases. At
$R=1.1$~\AA, the mean absolute errors of conventional QSCI at 200, 1000,
5000, and 20,000 shots are $31.15$, $14.68$, $4.63$, and $1.51\mHa$,
respectively. At the same shot counts, the errors are $32.58$, $13.09$,
$2.34$, and $0.71\mHa$ for SCI-NO RSU and $20.20$, $6.09$, $2.00$, and
$0.79\mHa$ for RES-NO RSU, respectively.
RES-NO gives the lowest mean error at the smaller shot counts,
whereas SCI-NO and RES-NO give similar mean errors at 20,000 shots.

At $R=1.8$~\AA, the mean absolute errors of conventional QSCI at 200, 1000,
5000, and 20,000 shots are $21.99$, $10.81$, $2.67$, and $0.68\mHa$,
respectively. At the same shot counts, the errors are $21.99$, $10.72$,
$1.25$, and $0.29\mHa$ for SCI-NO RSU and $20.92$, $2.96$, $0.52$, and
$0.12\mHa$ for RES-NO RSU, respectively.
RES-NO gives the lowest mean error at all tested shot counts, although the
three methods give similar mean errors at 200 shots.
At 20,000 shots, SCI-NO and RES-NO show smaller mean absolute errors than conventional
QSCI and retain more canonical Slater determinants in the final selected-CI space.
The numbers of retained determinants, averaged over eight measurement seeds,
are 205.5, 276.8, and 278.1 at $R=1.1$~\AA\ for conventional QSCI, SCI-NO RSU, and RES-NO RSU,
respectively. The corresponding averages at $R=1.8$~\AA\ are 229.8, 299.0, and 310.8.
This broader recovery of canonical determinants is consistent with the smaller selected-CI
energy errors obtained with the rotated orbital frames.

As in the ground-state calculations, measurements in the state-targeted
rotated frames generate canonical candidates that are difficult to observe
directly in conventional QSCI. These results show that SCI-NO and RES-NO
improve canonical determinant discovery at the same total number of
measurements.

\paragraph{Determinant discovery from the GQE-prepared state.}

We next repeat the excited-state comparison using an $S_1$ state prepared by GQE.
As described in Sec.~\ref{sec:reference-preparation}, the GQE circuit in the
present calculation is trained without quantum measurements of the
Hamiltonian, $\hat S^2$, or the overlap with the fixed ground state.
For each candidate circuit, Slater determinants are sampled by measuring the
prepared state in the computational basis.
The observed determinants are restricted to the target symmetry sector and used to
construct a finite spin-completed determinant space. The lowest singlet root in this
space subject to the fixed-ground orthogonality condition is obtained by a classical
selected-CI calculation, and its energy is used as the GQE reward. Thus, the quantum
contribution to the learning objective is limited to determinant sampling, while the
Hamiltonian, spin, and orthogonality conditions are evaluated classically in the
selected determinant space. The trained circuit is subsequently kept fixed for all
finite-shot comparisons.

Except for replacing the exact FCI sampling state with the GQE-prepared
state, these calculations follow the same determinant-sampling,
orbital-frame construction, finite-candidate selection, selected-CI, and
statistical procedures as the FCI-prepared calculations described above.
The final $S_1$ energy is obtained under the same fixed-ground
orthogonality condition in Eq.~\eqref{eq:fixed-ground-readout}. FCI
information is used only for post-computation assessment and is not used
for GQE training, orbital-frame construction, or determinant selection.

\begin{figure}[t]
\centering
\includegraphics[width=\linewidth]{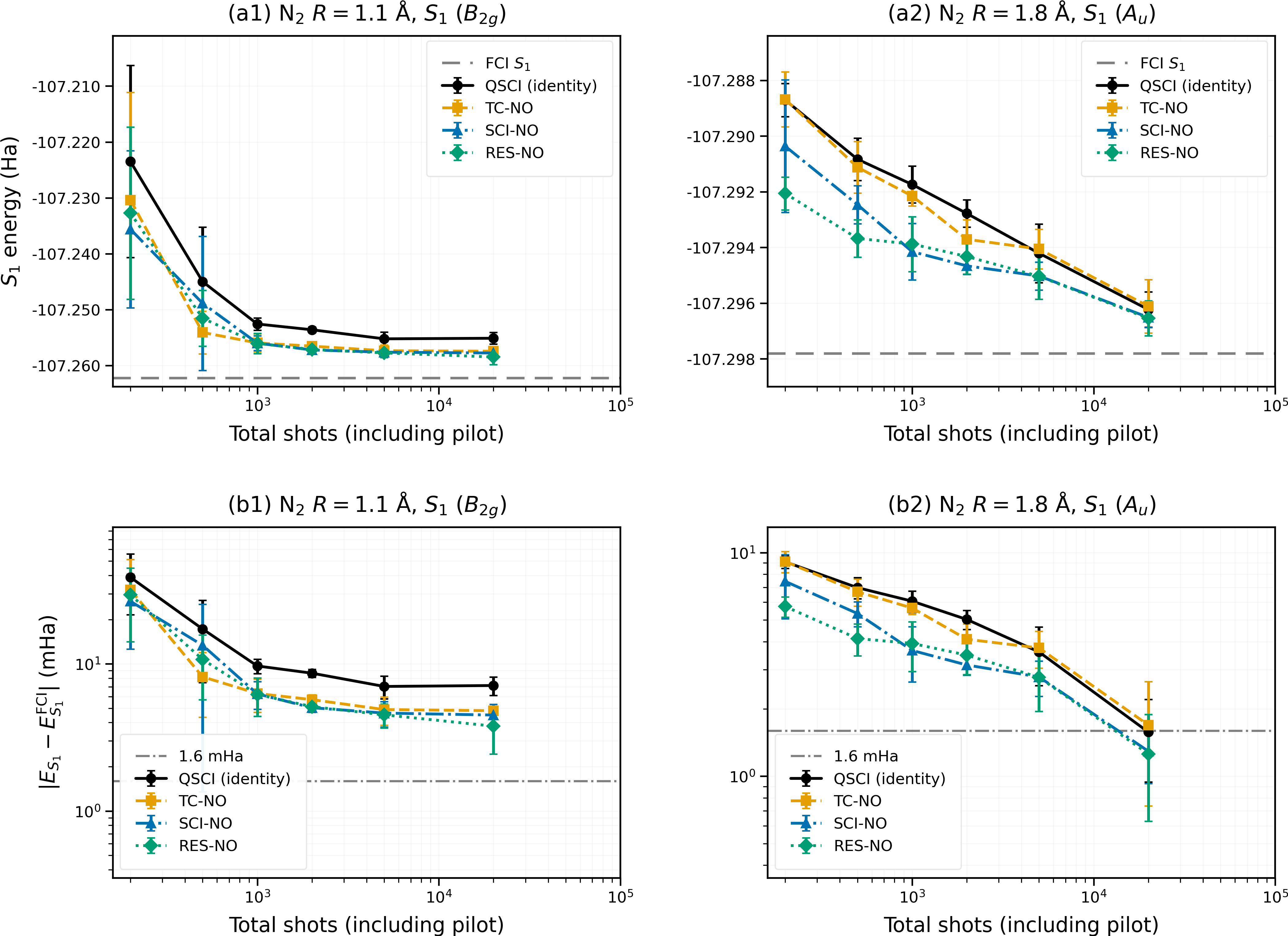}
\caption{Convergence of the $S_1$ energy obtained from the GQE circuit trained with
the sampled selected-CI reward at a target cap of 2000. The upper row shows the
selected-CI $S_1$ energy, and the dashed lines mark the active-space FCI energies.
The lower row shows the absolute error relative to FCI on a logarithmic scale, and
the dash-dotted line marks $1.6\mHa$. The left panels show N$_2$ at
$R=1.1$~\AA\ and the right panels show N$_2$ at $R=1.8$~\AA. QSCI samples
determinants in the canonical orbital basis. TC-NO, SCI-NO, and RES-NO use the
orbital frames defined in Sec.~\ref{par:tcno}. Symbols represent averages over eight
independent sampling runs, and error bars are sample standard deviations.}
\label{fig:rsu-s1-gqe}
\end{figure}

Figure~\ref{fig:rsu-s1-gqe} shows that the selected-CI energies approach the FCI
energies as the number of shots increases. At $R=1.1$~\AA, the mean error of
QSCI decreases from $38.77\mHa$ at 200 shots to $9.68\mHa$ at 1000 shots and
$7.13\mHa$ at 20,000 shots. At 200 shots, SCI-NO gives the largest reduction and
lowers the mean error to $26.64\mHa$. At 1000 shots, the errors are $9.68\mHa$ for
QSCI, $6.29\mHa$ for TC-NO, $6.25\mHa$ for SCI-NO, and $6.21\mHa$ for RES-NO.
At 20,000 shots, RES-NO gives the lowest mean error of $3.77\mHa$, followed by
SCI-NO at $4.49\mHa$, TC-NO at $4.81\mHa$, and QSCI at $7.13\mHa$.

At $R=1.8$~\AA, the mean error of QSCI decreases from $9.09\mHa$ at 200 shots to
$6.06\mHa$ at 1000 shots and $1.58\mHa$ at 20,000 shots. RES-NO gives errors of
$5.74$, $3.92$, and $1.26\mHa$ at the same shot counts, while SCI-NO gives
$7.44$, $3.65$, and $1.29\mHa$. At 20,000 shots, RES-NO gives the lowest mean
error, followed closely by SCI-NO; TC-NO gives $1.69\mHa$ and remains slightly
above QSCI. The state-targeted orbital frames therefore provide the clearest
improvement for the stretched geometry, where SCI-NO and RES-NO reduce the error
over the full range of shot counts tested.

These results demonstrate that a GQE circuit can be trained for an excited-state
QSCI calculation using a reward obtained from sampled determinants and a classical
selected-CI solve. The trained circuit supplies a determinant distribution that is
already effective for QSCI in the canonical orbital basis. Sampling in the
state-targeted orbital frames provides a further improvement by identifying additional
canonical determinants for the selected-CI calculation. The improvement depends on
the orbital frame and molecular geometry. RES-NO gives the lowest final mean error at
both bond lengths, with a particularly clear improvement at $R=1.8$~\AA.
The result should therefore be interpreted as a proof of concept for combining
sampling-based excited-state GQE preparation with QSCI and RSU, rather than as a
uniform advantage of one orbital frame under all conditions.

% ---------------------------------------------------------------

\subsection{Discussion}
\label{sec:discussion}
% ---------------------------------------------------------------
Using both FCI- and GQE-prepared states, we found that QSCI combined with
RSU recovers determinants that are difficult to discover by conventional
QSCI in the canonical orbital basis and thereby yields more accurate
energies with a finite number of shots. Sampling the same prepared state
in complementary orbital frames allows the state-targeted SCI-NO and
RES-NO frames to reveal canonical determinants that are rarely observed
by direct sampling in the canonical orbital basis. These additional
determinants enlarge the useful variational space available for the same
number of shots.
The strongest gains occur for N$_2$ at stretched bond lengths, where
strong static correlation distributes the wavefunction weight over a
broad set of determinants. One approximation in the present RSU procedure
is that the proposal score in Eq.~\eqref{eq:rsu-proposal-score} treats the
contributions from different observed rotated determinants independently
and therefore does not include interference between their back-transformed
amplitudes. This approximation may affect which canonical determinants are
selected from the finite candidate set.
The final accuracy depends both on the quality of the prepared state and
on whether its important determinant components are recovered with a
finite number of shots. RSU improves the latter by sampling the prepared
state in complementary orbital frames, but its effectiveness remains
limited when determinants important to the target state are insufficiently
represented in the prepared state.

When finite-shot sampling alone does not recover a sufficiently complete
determinant space, RSU could in principle be followed by a classical
configuration-expansion method such as Ext-SQD~\cite{barison2025quantum}.
Assessing whether this combination provides a practical advantage is left for
future study.
Similarly, RSU can be combined with the previously proposed wavefunction
refinement~\cite{kemmoku2026} (Supporting Information,
Sec.~\ref{si:wavefunction-refinement}). The refinement combines CI
wavefunctions obtained from separate calculations and reselects the
important determinants, whereas RSU expands determinant discovery by
sampling in different orbital frames. For the FCI-prepared ground state,
the combined procedure generally improved the energy when the size of the
final selected-CI space was restricted. This result shows that
rotated-frame sampling and wavefunction refinement can be used together
to improve the determinant content of a size-limited selected-CI space.
Detailed numerical comparisons are given in Supporting Information,
Sec.~\ref{si:rsu-refinement-combined} and
Fig.~\ref{fig:si-refinement-cap170}.
The orbital frames examined in this study are constructed classically and
require no additional quantum measurements of the one-particle reduced
density matrix. 

Although RSU recovers determinants that are rarely sampled in conventional QSCI (identity-frame QSCI), it requires, 
in addition to the state-preparation circuit, a frame-rotation circuit that applies the inverse 
orbital rotation to the prepared state before each rotated-frame measurement.
For a general 
real rotation of $n$ spatial orbitals, a Givens decomposition uses at most $n(n-1)/2$ two-orbital 
rotations per spin, with depth $O(n)$ on a linear qubit array~\cite{Kivlichan2018,Jiang2018}.
For the present $(10e,8o)$ space of the N$_{2}$ molecule, this general bound 
gives at most 56 Givens rotations, corresponding to 112 CNOT and 224 single-qubit rotations. 
The orbital transformations used in the present calculations are, however, much simpler 
than this general bound and require only a small number of nontrivial Givens rotations.
The typical RES-NO transformation requires two nontrivial Givens rotations
per spin, the TC-NO transformation requires one rotation in one spin, and
the SCI-NO transformation requires no more than six rotations in total in
the present calculations.
These results provide a concrete example in which improved determinant
discovery requires far fewer orbital rotations than the general upper
bound. With a suitable assignment of spin orbitals to qubits, this can
also substantially reduce the number of additional gates. 
The additional gate count depends on the assignment of spin-spatial orbitals to qubits and
whether the orbital pairs to be mixed can interact directly.
Gate errors can change the measured determinant probabilities and
hence the weights used for canonical determinant selection.
The shot reduction found in noiseless calculations should therefore
be considered together with the additional gates and their errors.
Detailed estimates are given in the Supporting Information.
The orbital frames examined here are guided by natural-orbital
constructions. Although they improve determinant discovery in the present
calculations, natural-orbital occupation is not necessarily the optimal
criterion for this purpose. Future work should therefore explore orbital
frames that more effectively recover important determinants while limiting
the number and depth of the additional orbital rotations, including their
efficient implementation for each molecular system.

While the present work was in progress, van Rossum et al. reported
measurements in multiple orbital bases constructed by classical NOCI
optimization~\cite{vanRossum2026}. Both NOCI-SQD and
RSU use orbital rotations to identify configurations that are rarely
observed when the prepared state is sampled only in the canonical orbital basis.
The main difference lies in how the sampled determinants are incorporated
into the final CI calculation. NOCI-SQD retains determinants expressed in
different orbital bases and solves a generalized eigenvalue problem using
their Hamiltonian and overlap matrices. RSU instead generates a finite set
of candidate determinants in a common canonical basis from the observed
rotated determinants. The candidates are ranked using squared expansion
coefficients weighted by the observed counts. The final Hamiltonian is
then diagonalized in an orthonormal selected-CI space, without an overlap
matrix between rotated determinants.
The orbital frames are also constructed differently. NOCI-SQD optimizes
orbital rotations and CI coefficients to lower the energy of a
nonorthogonal expansion of rotated Hartree--Fock determinants. In the
present work, TC-NO is constructed from the known state-preparation
circuit, while SCI-NO and RES-NO use the one-particle density matrices of
an initial selected-CI wavefunction and its external residual,
respectively.
Whereas the reported NOCI-SQD benchmarks concern ground
states~\cite{vanRossum2026}, we examine RSU for both ground and target
excited states. For the latter, GQE is adapted to prepare sampling states
specifically for the target excited state.
Despite these differences, both approaches exploit the orbital-rotation freedom
of Slater determinants, described by Thouless transformations, to improve the
discovery of useful configurations from finite measurements.
Together, they motivate the use of orbital rotations not only to represent
electronic wavefunctions, but also to make their measurement more informative
for selected CI.

The present results provide a proof of concept for combining state-specific
GQE preparation with RSU to improve determinant discovery for ground- and
excited-state QSCI. Future work will examine the effects of noise and
validate the approach on quantum hardware, assessing whether the sampling
improvements persist in the presence of state-preparation and
orbital-rotation errors.

\section*{Acknowledgments}
Y.O. acknowledges the support of Dr. H. Noda at FUJIFILM Corporation and thanks Satoru Shoji and Rahul Bhowmick of FUJITSU Limited for carefully reading the manuscript. 
The author used mainly Claude Opus 4.8 and GPT-6.1 Sol
for implementing and validating the algorithms.
We also used GPT-6 Luna during manuscript preparation to facilitate
the writing process. All computational results and scientific conclusions were verified by the author.

\section*{Corresponding Author}
Yukihiro Okuno -- FUJIFILM Corporation, Minamiashigara, Kanagawa 250-0193, Japan;
\href{https://orcid.org/0000-0002-6114-5129}{orcid.org/0000-0002-6114-5129};
Email: \href{mailto:yukihiro.okuno@fujifilm.com}{yukihiro.okuno@fujifilm.com}.

\section*{Competing interests}
The author declares no financial or non-financial competing interests.

\section*{Code and data availability statement}
All codes and data are available upon reasonable request.

% ======================================================================
\section*{Supporting Information}
The Supporting Information contains interference in RSU proposal scores and
coherent likelihood tests, wavefunction refinement and its combination with RSU,
ground-state GQE support growth, and the classical construction and resource
limitations of SCI-NO, RES-NO, and RSU unrotation.
\bibliography{references}
% ======================================================================

% ======================================================================
% Combined Supporting Information
% ======================================================================
\clearpage
\setcounter{section}{0}
\setcounter{subsection}{0}
\setcounter{subsubsection}{0}
\setcounter{figure}{0}
\setcounter{table}{0}
\setcounter{equation}{0}
\renewcommand{\thesection}{S\arabic{section}}
\renewcommand{\thefigure}{S\arabic{figure}}
\renewcommand{\thetable}{S\arabic{table}}
\renewcommand{\theequation}{S\arabic{equation}}
\renewcommand{\theHsection}{SI.\arabic{section}}
\renewcommand{\theHsubsection}{SI.\arabic{section}.\arabic{subsection}}
\renewcommand{\theHsubsubsection}{SI.\arabic{section}.\arabic{subsection}.\arabic{subsubsection}}
\renewcommand{\theHfigure}{SI.\arabic{figure}}
\renewcommand{\theHtable}{SI.\arabic{table}}
\renewcommand{\theHequation}{SI.\arabic{equation}}
\begin{center}
{\LARGE\bfseries Supporting Information:\\[0.5em]
Rotated-Frame Determinant Discovery for Ground- and Excited-State Quantum Selected Configuration Interaction\par}
\vspace{1.5em}
{\large Yukihiro Okuno\par}
\vspace{0.4em}
{\large FUJIFILM Corporation, Minamiashigara, Kanagawa 250-0193, Japan\par}
\vspace{1em}
{\large \today\par}
\end{center}
\vspace{2em}
\section{Construction of orbital frames for rotated-determinant sampling}
\label{si:frame-construction}
\label{si:classical-resources}
\label{si:frame-cost}
This section describes how the TC-NO, SCI-NO, and RES-NO orbital frames are
constructed. Each frame is obtained by diagonalizing a spin-resolved
one-particle reduced density matrix (1-RDM), and its eigenvectors define the
orbitals used for the rotated-frame measurements. The three frames differ
only in the 1-RDM used. TC-NO uses an approximate 1-RDM constructed from the
excitation operators and signed rotation angles of the GQE circuit. SCI-NO
uses the 1-RDM of an initial selected-CI wavefunction obtained from
measurements in the canonical orbital basis. RES-NO uses the 1-RDM of the
normalized residual of that wavefunction outside the initial selected-CI
space. All three constructions are classical and require neither a quantum
measurement of a 1-RDM nor a Hamiltonian diagonalization outside the initial
selected-CI space. The generation of canonical candidates from the
rotated-frame measurements is described in Sec.~\ref{si:candidate-backtransform}.

% ---------------------------------------------------------------------
\subsection{Orbital rotations and one-particle density matrices}
\label{si:frame-common}
We consider an active space of $n_{\mathrm{orb}}$ canonical (Hartree--Fock)
spatial orbitals $\{\phi_p\}$ containing $N_\alpha$ $\alpha$ electrons and
$N_\beta$ $\beta$ electrons. Canonical determinants are denoted by $\ket{I}$,
$\ket{J}$, and $\ket{A}$. For each spin $\sigma\in\{\alpha,\beta\}$, the rotated
orbitals are defined by a real orthogonal
$n_{\mathrm{orb}}\times n_{\mathrm{orb}}$ matrix $R_\sigma$, whose columns
contain their expansion coefficients in the canonical orbitals:
\begin{equation}
 \phi^{F}_{p\sigma}=\sum_{q}\phi_{q\sigma}(R_\sigma)_{qp},
 \qquad R_\sigma^{T}R_\sigma=I .
 \label{eq:si-frame-orbitals}
\end{equation}
The corresponding creation operators transform as
\begin{equation}
 \hat b^{\dagger}_{p\sigma}=\sum_{q}\hat a^{\dagger}_{q\sigma}(R_\sigma)_{qp},
 \label{eq:si-frame-operators}
\end{equation}
and an orbital frame is the pair $F=(R_\alpha,R_\beta)$. The rotated
determinant $\ket{\varphi_{F,d}}$ is constructed from the rotated orbitals
with the occupation string $d$. Its coefficient on each canonical determinant
is the product of the $\alpha$ and $\beta$ minors of $R_\alpha$ and $R_\beta$
formed by the occupied orbitals of the two determinants, as used for the
transformation back to the canonical orbital basis in the main text
(see also Eq.~\eqref{eq:si-candidate-component}). The identity frame,
$R_\alpha=R_\beta=I$, reproduces conventional QSCI sampling.

The spin-resolved one-particle reduced density matrix (1-RDM) of a
normalized state $\ket{\Psi}$ is
\begin{equation}
 \gamma^{\sigma}_{pq}=\bra{\Psi}\hat a^\dagger_{p\sigma}\hat a_{q\sigma}\ket{\Psi},
 \qquad \operatorname{Tr}\gamma^\sigma=N_\sigma .
 \label{eq:si-frame-rdm-general}
\end{equation}
If $\ket{\Psi}=\sum_I C_I\ket{I}$ is given as a CI expansion, the 1-RDM is
assembled from one-body coupling coefficients between pairs of determinants,
\begin{equation}
 \gamma^{\sigma}_{pq}=\sum_{I,J}C_I^{*}C_J
 \bra{I}\hat a^\dagger_{p\sigma}\hat a_{q\sigma}\ket{J}.
 \label{eq:si-frame-rdm-ci}
\end{equation}
The diagonal elements of the 1-RDM are obtained from the orbital occupations of each determinant.
The off-diagonal elements arise from pairs of determinants that differ only by moving
one electron of spin $\sigma$ from orbital $q$ to orbital $p$.
Determinants that differ by two or more orbital substitutions do not contribute to the 1-RDM.
The natural orbitals of the 
state are the eigenvectors of the 1-RDM,
\begin{equation}
 \gamma^{\sigma}R_\sigma=R_\sigma\Lambda_\sigma,
 \qquad \Lambda_\sigma=\operatorname{diag}(\lambda_{1\sigma},\dots,
 \lambda_{n_{\mathrm{orb}}\sigma}),
 \label{eq:si-frame-no-general}
\end{equation}
where the eigenvalues $\lambda_{p\sigma}$ are the natural occupation numbers.
Each orbital rotation below is obtained from Eq.~\eqref{eq:si-frame-no-general} by a
different choice of $\gamma^\sigma$. The eigenvalue problem has the dimension
$n_{\mathrm{orb}}$ of the orbital space, not the dimension of the determinant
space. For the real wavefunctions considered here all matrices are real, and
the eigenvectors are chosen with a fixed sign convention. Changing the sign of
an eigenvector changes only the overall sign of the rotated determinants that
contain that orbital and does not change any measured probability.

% ---------------------------------------------------------------------
\subsection{Token-cumulant natural-orbital frame (TC-NO)}
\label{si:tcno-construction}
TC-NO uses the excitation operators and signed coefficients in the GQE
preparation circuit. For a single-determinant reference, these quantities give
approximate single- and double-excitation amplitudes. The amplitudes are used
to construct the approximate 1-RDM in Sec.~\ref{si:tcno-single}.
For a multiconfigurational reference, the 1-RDM is instead evaluated with the
first-order commutator expression in Sec.~\ref{si:tcno-seed}.
The ground-state circuits start from the Hartree--Fock determinant and use
Sec.~\ref{si:tcno-single}, whereas the excited-state circuits
start from a configuration state function and use Sec.~\ref{si:tcno-seed}.
These matrices are used only to obtain orbital eigenvectors. They are not
the exact 1-RDMs of the complete GQE-prepared wavefunctions.

\subsubsection{Single- and double-excitation amplitudes from the GQE circuit}

In GQE, each token of the circuit sequence selects one excitation operator from the operator pool
together with a signed rotation angle. For token $\ell$ of state $k$, we denote the excitation
label by $\mu_{k,\ell}$ and the signed rotation angle by $w_{k,\ell}\in\mathcal W$, where $\mathcal W$
is the discrete angle set of main-text Eq.~\eqref{eq:tcno-token-weight}.
The corresponding anti-Hermitian generator is
\begin{equation}
  \hat G_\mu=\hat E_\mu-\hat E_\mu^\dagger,
  \label{eq:si-tcno-token-weight}
\end{equation}
where $\hat E_\mu$ is the excitation operator.

The ordered circuit and its expansion about the reference determinant are
\begin{equation}
 U(\boldsymbol\tau_k)=e^{w_{k,L_k}\hat G_{\mu_{k,L_k}}}
 \cdots e^{w_{k,1}\hat G_{\mu_{k,1}}},
 \label{eq:si-tcno-circuit}
\end{equation}
\begin{equation}
 U(\boldsymbol\tau_k)\ket{\Phi_k}
 =\ket{\Phi_k}+\sum_\ell w_{k,\ell}
 \hat E_{\mu_{k,\ell}}\ket{\Phi_k}+O(w^2).
 \label{eq:si-tcno-leading-state}
\end{equation}
Here the reference $\ket{\Phi_k}$ is a closed-shell determinant, and the
excitation labels refer to its occupied and virtual orbitals. Identity
tokens have zero angle. Because Eq.~\eqref{eq:si-tcno-leading-state} is
linear in the angles, repeated tokens for the same excitation contribute
additively to its leading-order amplitude, and tokens with opposite signs
cancel at this order.
We use $\sigma\in\{\alpha,\beta\}$ for spin and $\bar\sigma$ for the opposite
spin. Indices $i,j,l$ refer to spatial orbitals occupied in $\ket{\Phi_k}$,
and $a,b,c$ to those unoccupied in $\ket{\Phi_k}$. The accumulated single-excitation
amplitudes are
\begin{equation}
  s^{(k),\sigma}_{ia} = \sum_{\ell:\mu_{k,\ell}=(i\sigma\rightarrow a\sigma)} w_{k,\ell}.
  \label{eq:si-tcno-singles}
\end{equation}
Similarly, the opposite-spin double-excitation amplitudes
are obtained by summing the signed weights of tokens
corresponding to the same excitation:
\begin{equation} T^{(k)}_{ia,jb} = \sum_{\ell:\mu_{k,\ell} =((i\alpha,j\beta)\rightarrow(a\alpha,b\beta))} w_{k,\ell}.
  \label{eq:si-tcno-opposite-doubles}
\end{equation}
The first index pair of $T^{(k)}$ always belongs to the $\alpha$ spin and the
second to the $\beta$ spin. To write the density-matrix expressions uniformly
for either spin, set
\begin{equation}
  D^{(k),\alpha\beta}_{ia,jb}=T^{(k)}_{ia,jb}, \qquad D^{(k),\beta\alpha}_{ia,jb}=T^{(k)}_{jb,ia}.
  \label{eq:si-tcno-opposite-order}
\end{equation}
For a same-spin double excitation, the operator pool uses the representative
ordering $i<j$ and $a<b$. We sum the weights of tokens with those indices to
obtain $D^{(k),\sigma\sigma}_{ij,ab}$ and extend the result by fermionic
antisymmetry,
\begin{equation}
  D^{(k),\sigma\sigma}_{ji,ab} =-D^{(k),\sigma\sigma}_{ij,ab}, \qquad D^{(k),\sigma\sigma}_{ij,ba} =-D^{(k),\sigma\sigma}_{ij,ab}.
  \label{eq:si-tcno-same-spin}
\end{equation}
Thus, exchanging both the occupied and the virtual indices leaves the
amplitude unchanged, and $D^{(k),\sigma\sigma}_{ii,ab}=D^{(k),\sigma\sigma}_{ij,aa}=0$.
With these amplitudes, the leading-order correlated state can be written in
the familiar CI-like form
\begin{equation}
 \ket{\Psi_k}\approx\ket{\Phi_k}
 +\sum_{\sigma}\sum_{ia}s^{(k),\sigma}_{ia}\ket{\Phi^{a\sigma}_{i\sigma}}
 +\sum_{ia,jb}T^{(k)}_{ia,jb}\ket{\Phi^{a\alpha b\beta}_{i\alpha j\beta}}
 +\frac14\sum_{\sigma}\sum_{ij,ab}D^{(k),\sigma\sigma}_{ij,ab}
 \ket{\Phi^{a\sigma b\sigma}_{i\sigma j\sigma}},
 \label{eq:si-tcno-ci-form}
\end{equation}
where $\ket{\Phi^{a\sigma}_{i\sigma}}$ and the doubly substituted determinants
are defined with the usual phase convention of second quantization. The
factor $1/4$ compensates for the unrestricted sums over the two pairs
$(i,j)$ and $(a,b)$.

\subsubsection{Approximate 1-RDM for a single-determinant reference}
\label{si:tcno-single}
For each state and spin, these amplitudes define a real approximate 1-RDM
$\tilde\gamma^{(k),\sigma}$. The reference--single transition contributes
linearly to the occupied--virtual block,
\begin{equation}
  \tilde{\gamma}^{(k),\sigma}_{ia} = \tilde{\gamma}^{(k),\sigma}_{ai} = s^{(k),\sigma}_{ia}.
  \label{eq:si-tcno-ov}
\end{equation}
whereas doubly excited determinants are not connected to the reference by a
one-body operator and contribute only quadratically to the
virtual--virtual and occupied--occupied blocks:
\begin{equation}
  \begin{split} \tilde{\gamma}^{(k),\sigma}_{ab} ={}& \sum_i s^{(k),\sigma}_{ia}s^{(k),\sigma}_{ib} +\frac{1}{2}\sum_{ijc} D^{(k),\sigma\sigma}_{ij,ac} D^{(k),\sigma\sigma}_{ij,bc}\\ &+ \sum_{i\bar{j}\bar{c}} D^{(k),\sigma\bar{\sigma}}_{ia,\bar{j}\bar{c}} D^{(k),\sigma\bar{\sigma}}_{ib,\bar{j}\bar{c}}, \end{split}
  \label{eq:si-tcno-vv}
\end{equation}
and
\begin{equation}
  \begin{split} \tilde{\gamma}^{(k),\sigma}_{ij} ={}& \delta_{ij} -\sum_a s^{(k),\sigma}_{ia}s^{(k),\sigma}_{ja} -\frac{1}{2}\sum_{lab} D^{(k),\sigma\sigma}_{il,ab} D^{(k),\sigma\sigma}_{jl,ab}\\ &- \sum_{a\bar{l}\bar{b}} D^{(k),\sigma\bar{\sigma}}_{ia,\bar{l}\bar{b}} D^{(k),\sigma\bar{\sigma}}_{ja,\bar{l}\bar{b}}. \end{split}
  \label{eq:si-tcno-oo}
\end{equation}
Here $\delta_{ij}$ is the Kronecker delta, barred orbital indices belong to
the opposite spin, and all sums run over the occupied or virtual ranges
indicated above. The factors of $1/2$ in the same-spin double-excitation
terms of Eqs.~\eqref{eq:si-tcno-vv} and \eqref{eq:si-tcno-oo} correct for the
double counting of the occupied pair $(i,j)$ and of the virtual pair $(a,b)$,
respectively. The positive virtual--virtual contractions describe the
occupation of orbitals into which electrons are promoted, and the
corresponding negative occupied--occupied contractions describe the holes
left behind. In each spin, the gain in virtual occupation equals the loss in
occupied occupation, so that
\begin{equation}
 \sum_a\tilde\gamma^{(k),\sigma}_{aa}
 =\sum_i\left(1-\tilde\gamma^{(k),\sigma}_{ii}\right),
 \qquad
 \operatorname{Tr}\tilde\gamma^{(k),\sigma}=N_\sigma .
 \label{eq:si-tcno-trace}
\end{equation}
Quadratic single--double contributions to the occupied--virtual block,
corrections arising from the ordering of the exponentials in
Eq.~\eqref{eq:si-tcno-circuit}, and higher-order terms are omitted, and the
normalization of the state in Eq.~\eqref{eq:si-tcno-ci-form} is not imposed.
As a result, the eigenvalues of $\tilde\gamma^{(k),\sigma}$ need not lie in
the physical occupation interval $[0,1]$. This is acceptable because the
matrix is used only to propose orbital frames, not to compute properties.

Each matrix is symmetrized to remove numerical asymmetry and diagonalized,
\begin{equation}
  G^{(k),\sigma} = \frac{1}{2} ( \tilde{\gamma}^{(k),\sigma} + (\tilde{\gamma}^{(k),\sigma})^{T} ), \qquad G^{(k),\sigma}R^{(k)}_\sigma = R^{(k)}_\sigma\Lambda^{(k)}_\sigma.
  \label{eq:si-tcno-frame}
\end{equation}
The columns of $R^{(k)}_\sigma$ are the orbital eigenvectors, ordered by
decreasing eigenvalue, and $\Lambda^{(k)}_\sigma$ is the diagonal matrix of
the corresponding approximate occupations. Orbital phases are fixed and each
rotation is chosen to have determinant $+1$. The state-specific TC-NO frame
is $F_k=(R^{(k)}_\alpha,R^{(k)}_\beta)$. If all amplitudes vanish,
$G^{(k),\sigma}$ is diagonal in the canonical orbitals and the TC-NO frame
reduces to the canonical orbital basis.

\subsubsection{Orbitals from differences between density matrices}
In multi-state calculations, frames constructed from density differences can
complement the state-specific frames. For example, a frame describing the
change in density between states 0 and 1 is obtained by diagonalizing
\begin{equation}
  \Delta G^\sigma = G^{(1),\sigma}-G^{(0),\sigma},
  \label{eq:si-tcno-difference}
\end{equation}
with the eigenvectors ordered by the magnitude of the difference
eigenvalues. Orbitals with large positive or negative eigenvalues are those
whose occupation changes most upon excitation.
The resulting orbital rotations are physically motivated candidates; they are not constructed to
minimize or maximize the sampling entropy in main-text Eq.~\eqref{eq:rsu-entropy}.
This construction is not used in the calculations reported here.

\subsubsection{TC-NO for a multiconfigurational reference}
\label{si:tcno-seed}
For the excited-state GQE circuit, the starting state is a spin-adapted
configuration state function (main-text Sec.~\ref{sec:reference-preparation}), that is,
a normalized linear combination of orthonormal determinants. The single-determinant expressions
above therefore do not apply, and the 1-RDM is evaluated directly.
Let $\ket{\chi(c)}=\sum_j c_j\ket{\chi_j}$ denote this starting state ($\ket{\Psi_k^{\mathrm{seed}}}$
in the main text), and let $U=e^{\theta_L\hat G_L}\cdots e^{\theta_1\hat G_1}$ be the GQE circuit,
with $\theta_\ell=w_{k,\ell}$ and $\hat G_\ell=\hat G_{\mu_{k,\ell}}$.
To first order in the rotation angles, the transformed density operator satisfies
\begin{equation}
 U^\dagger\hat O_{pq\sigma}U
 =\hat O_{pq\sigma}+\sum_{\ell=1}^{L}\theta_\ell
 [\hat O_{pq\sigma},G_\ell]+O(\theta^2),
 \label{eq:si-seed-tcno-heisenberg}
\end{equation}
so that the spin-resolved 1-RDM is, to the same order,
\begin{equation}
 \widetilde\gamma^\sigma_{pq}
 =\langle\chi(c)|O_{pq\sigma}|\chi(c)\rangle
 +\sum_{\ell=1}^L\theta_\ell
 \langle\chi(c)|[O_{pq\sigma},G_\ell]|\chi(c)\rangle,
 \qquad O_{pq\sigma}=a^\dagger_{p\sigma}a_{q\sigma}.
 \label{eq:si-seed-tcno-density}
\end{equation}
Writing the seed expectation values explicitly,
\begin{equation}
 \langle\chi(c)|X|\chi(c)\rangle
 =\sum_{j}|c_j|^2\langle\chi_j|X|\chi_j\rangle
 +\sum_{j\neq j'}c_j^{*}c_{j'}\langle\chi_j|X|\chi_{j'}\rangle,
 \label{eq:si-seed-tcno-transition}
\end{equation}
for $X=O_{pq\sigma}$ or $X=[O_{pq\sigma},G_\ell]$, shows that this expression
includes transition-density contributions between different seed states.
The generator and rotation angle of each exponential are taken from the GQE circuit.
The orbital eigenvectors of the symmetric part of
$\widetilde\gamma^\sigma$ define the corresponding TC-NO orbital rotation. This
first-order density is used only for orbital construction and is not the exact
1-RDM of $U\ket{\chi(c)}$.

% ---------------------------------------------------------------------
\subsection{Selected-CI natural-orbital frame (SCI-NO)}
\label{si:scino-explicit}
SCI-NO is constructed from an initial selected-CI wavefunction. For
electronic state $k$, canonical determinants are selected from the initial
measurements in the canonical orbital basis, and the determinants required
for spin completion are added. We denote the resulting set by $B$, which
corresponds to $\mathcal B_0$ in the main text. The Hamiltonian is
diagonalized in $B$ within the target spatial symmetry and the singlet
subspace. For the ground state, the lowest root is used. For the target
excited state, the lowest root orthogonal to the fixed ground-state
approximation $\ket{g}$ is used, as in main-text
Eq.~\eqref{eq:fixed-ground-readout}. The resulting normalized CI
wavefunction and its energy are
\begin{equation}
 \ket{\psi_k^{(0)}}=\sum_{I\in B}c_I^{(k)}\ket{I},
 \qquad \sum_{I\in B}|c_I^{(k)}|^2=1,
 \qquad E_k^{(0)}=\bra{\psi_k^{(0)}}\hat H\ket{\psi_k^{(0)}}.
 \label{eq:si-scino-initial}
\end{equation}

The spin-resolved 1-RDM of the initial selected-CI wavefunction is
\begin{equation}
 \gamma_{k,pq}^{\sigma}
 =\sum_{I,J\in B}(c_I^{(k)})^*c_J^{(k)}
 \bra{I}\hat a^\dagger_{p\sigma}\hat a_{q\sigma}\ket{J}.
 \label{eq:si-scino-rdm}
\end{equation}
Only pairs of determinants that are identical or differ by one spin-orbital
substitution contribute. For a fixed ordering of spin orbitals, the
one-body coupling coefficients are evaluated as follows.
Let $n_t(J)$ denote the occupation of spin orbital $t$ in determinant $J$.
Removing an electron from the occupied spin orbital $q\sigma$ gives $J^-$.
Placing this electron in the unoccupied spin orbital $p\sigma$ gives $J'$.
Then
\begin{equation}
 \bra{I}\hat a^\dagger_{p\sigma}\hat a_{q\sigma}\ket{J}
 =(-1)^{\sum_{t<q\sigma}n_t(J)+\sum_{t<p\sigma}n_t(J^-)}
 \delta_{I,J'}.
 \label{eq:si-onebody-sign}
\end{equation}
The coefficient is zero when either occupation condition fails. Here the
inequalities refer to positions in the fixed spin-orbital ordering. For $p=q$
the expression reduces to $n_{p\sigma}(J)\delta_{IJ}$. Thus the density
matrix is accumulated from pairs of initial selected determinants that are identical or
differ by one orbital substitution, and no wavefunction outside $B$ is
needed. The resulting matrix satisfies
\begin{equation}
 (\gamma_k^\sigma)^\dagger=\gamma_k^\sigma,
 \qquad \operatorname{Tr}\gamma_k^\sigma=N_\sigma,
 \qquad 0\le\lambda^{(S)}_{k,p\sigma}\le1,
 \label{eq:si-scino-checks}
\end{equation}
where the last relation follows from the Pauli principle for the normalized
wavefunction $\ket{\psi_k^{(0)}}$. The SCI-NO orbitals and frame are defined by
\begin{equation}
 \gamma_k^\sigma R_{k,\sigma}^{(S)}
 =R_{k,\sigma}^{(S)}\Lambda_{k,\sigma}^{(S)},
 \qquad (R_{k,\sigma}^{(S)})^\dagger R_{k,\sigma}^{(S)}=I,
 \label{eq:si-scino-eigenproblem}
\end{equation}
\begin{equation}
 F_k^{(S)}=(R_{k,\alpha}^{(S)},R_{k,\beta}^{(S)}),
 \label{eq:si-scino-frame}
\end{equation}
with $\Lambda^{(S)}_{k,\sigma}=\operatorname{diag}(\lambda^{(S)}_{k,1\sigma},
\dots,\lambda^{(S)}_{k,n_{\mathrm{orb}}\sigma})$. The superscript $(S)$
denotes SCI-NO. Each density matrix has the dimension of the active
spatial-orbital space. For the real wavefunctions considered here, the
rotations are orthogonal. For a state of definite symmetry in the Abelian
point group used here, the 1-RDM has no elements between orbitals of
different irreducible representations, and diagonalization is performed
within each irreducible-representation block so that accidental degeneracies
do not mix different symmetry blocks. Eigenvectors are ordered by decreasing
occupation within each block. Phases and the basis within any degenerate
subspace are fixed by a consistent convention. These choices define only the
orbitals of the frame; they involve neither additional measurements nor a
Hamiltonian eigenvalue problem in determinant space.

% ---------------------------------------------------------------------
\subsection{Residual natural-orbital frame (RES-NO)}
\label{si:resno-explicit}
RES-NO is constructed from the part of $(\hat H-E_k^{(0)})\ket{\psi_k^{(0)}}$ that lies
outside the initial selected-CI space. This residual is calculated from the initial
CI wavefunction and the molecular integrals. Its spin-resolved 1-RDM is then diagonalized
to obtain the RES-NO orbitals. All these calculations are performed classically.

\subsubsection{Residual outside the selected determinant space}
For electronic state $k$, let $B$ denote the canonical determinants included in
the initial selected-CI calculation. We also define $\mathcal X_k$ as the set of all
canonical determinants with the same spatial symmetry as state $k$.
The corresponding projection operators are
\begin{equation}
  P_B=\sum_{I\in B}\ket{I}\bra{I},\qquad Q_B=I-P_B,\qquad Q_k=\sum_{A\in\mathcal X_k}\ket{A}\bra{A}.
  \label{eq:si-resno-projectors}
\end{equation}
Here, $P_B$ retains the determinants of the initial selected-CI space, and
$Q_B$ removes them. The operator $Q_k$ retains only determinants with the
spatial symmetry of the target state. Because $P_B$ and $Q_k$ are both
diagonal in the canonical determinant basis, they commute, and
$Q=Q_BQ_k$ selects determinants with the target symmetry outside $B$. We
define
\begin{equation}
 Q=Q_BQ_k,\qquad
 \ket{u_k}=Q(\hat H-E_k^{(0)})\ket{\psi_k^{(0)}},
 \qquad \ket{v}=Q\ket{g}.
 \label{eq:si-resno-unprojected}
\end{equation}
Here $\ket{u_k}$ is the external residual in the target symmetry, and
$\ket{v}$ is the component of the fixed ground state in the same external
space. For an excited state, the residual is further made orthogonal to
$\ket{g}$. When $\langle v|v\rangle>0$, the component along $\ket{v}$ is
removed:
\begin{equation}
  Q_g=I-\frac{\ket{v}\bra{v}}{\langle v|v\rangle}, \qquad \ket{r_k}=Q_g\ket{u_k}.
  \label{eq:si-resno-ground-projector}
\end{equation}
If $\ket{v}=0$, we set $Q_g=I$. The operator $Q_g$ corresponds to
$Q_{g,\mathrm{out}}$ in the main text. This projection is not applied in
the ground-state calculation.
The resulting residual satisfies
\begin{equation}
  P_B\ket{r_k}=0,\qquad Q_k\ket{r_k}=\ket{r_k}.
  \label{eq:si-resno-space-constraints}
\end{equation}
For an excited-state calculation with the fixed-ground projection, it
also satisfies
\begin{equation}
  \braket{g}{r_k}=0.
  \label{eq:si-resno-ground-constraint}
\end{equation}

When the initial selected-CI wavefunction is obtained without an active
orthogonality constraint, it is an eigenvector of the Hamiltonian in $B$ and
therefore satisfies
\begin{equation}
  P_B(\hat H-E_k^{(0)})\ket{\psi_k^{(0)}}=0.
  \label{eq:si-resno-internal-residual}
\end{equation}
In this case, $Q_B$ does not change the residual in exact arithmetic.
When the fixed-ground constraint is active, the left-hand side is generally
nonzero along the projection of $\ket{g}$ onto $B$, and $Q_B$ removes this
component. In the $S_1$ calculations reported here, the target state
and $\ket{g}$ have different spatial symmetries, so the constraint is inactive.
In both cases, $Q_B$ ensures that RES-NO is constructed only from determinants outside
the initial selected-CI space.

\subsubsection{Evaluation of residual coefficients}

For a canonical determinant $A\in\mathcal X_k$ outside $B$, its coefficient
in the symmetry-selected external residual $\ket{u_k}$ is
\begin{equation}
 u_{k,A}
 =\sum_{I\in B}H_{AI}c_I^{(k)},
 \qquad
 H_{AI}=\bra{A}\hat H\ket{I}.
 \label{eq:si-residual-coefficient}
\end{equation}
The term $-E_k^{(0)}\ket{\psi_k^{(0)}}$ does not contribute because
$\ket{\psi_k^{(0)}}$ contains only determinants in $B$. Since the electronic
Hamiltonian contains one- and two-electron operators, $H_{AI}$ can be nonzero
only when $A$ and $I$ differ by one or two spin-orbital substitutions.

Using the antisymmetrized two-electron integrals
\begin{equation}
 \langle pq\Vert rs\rangle
 =\langle pq|rs\rangle-\langle pq|sr\rangle,
 \label{eq:si-antisymmetrized-integrals}
\end{equation}
where $\langle pq|rs\rangle=(pr|qs)$ in chemists' notation,
the Slater--Condon matrix element for a single substitution is
\begin{equation}
 H_{AI}
 =\Gamma_{AI}
 \left(h_{ai}+\sum_{m\in I}\langle am\Vert im\rangle\right),
 \qquad
 \ket{A}=\hat a^\dagger_a\hat a_i\ket{I},
 \label{eq:si-slater-condon-single}
\end{equation}
and that for a double substitution is
\begin{equation}
 H_{AI}
 =\Gamma_{AI}\langle ab\Vert ij\rangle,
 \qquad
 \ket{A}=\hat a^\dagger_a\hat a^\dagger_b
 \hat a_j\hat a_i\ket{I}.
 \label{eq:si-slater-condon-double}
\end{equation}
Here $i$ and $j$ are occupied spin orbitals in $I$, while $a$ and $b$ are
unoccupied spin orbitals. The sum in
Eq.~\eqref{eq:si-slater-condon-single} runs over the spin orbitals occupied in
$I$. The factor $\Gamma_{AI}=\pm1$ is the sign obtained when the spin-orbital
occupations are restored to the fixed ordering used for the determinants.

The external determinants that can have nonzero coefficients form the set
\begin{equation}
 C_k=\left\{A\in\mathcal X_k\setminus B\ \middle|\
 H_{AI}\ne0\text{ for at least one }I\in B\right\}.
 \label{eq:si-connected-set}
\end{equation}
In practice, $C_k$ is generated directly from the
determinants in $B$: for each $I\in B$, all single
and double substitutions are formed, and those that
have the target spatial symmetry and are not
contained in $B$ are kept.
Neither $\mathcal X_k$ nor the full determinant
space is enumerated.
Each external coefficient in Eq.~\eqref{eq:si-residual-coefficient} sums,
with fermionic signs, the couplings of $A$ to all determinants in $B$. The
calculation therefore amounts to a single application of the Hamiltonian to
$\ket{\psi_k^{(0)}}$; no Hamiltonian diagonalization is performed in the
external space.

For a ground-state calculation, the coefficients $u_{k,A}$ are the final
residual coefficients. For an excited-state calculation, the component
parallel to the fixed ground-state wavefunction is removed. With
$v_A=\braket{A}{v}$, the projected residual is
\begin{equation}
 \ket{r_k}
 =\ket{u_k}-\ket{v}\frac{\langle v|u_k\rangle}{\langle v|v\rangle},
 \qquad
 r_{k,A}
 =u_{k,A}-v_A
 \frac{\sum_{A'}v_{A'}^*u_{k,A'}}{\sum_{A'}|v_{A'}|^2},
 \label{eq:si-ground-residual-projection}
\end{equation}
when $\langle v|v\rangle>0$. If $\ket{v}=0$, then
$\ket{r_k}=\ket{u_k}$.
We denote the support of the residual by $D_k$. For the ground state,
$D_k=C_k$. For an excited state, when the fixed ground state is stored as a
sparse CI expansion, $D_k$ is the union of $C_k$ and the external
determinants of $\ket{g}$ in $\mathcal X_k$, and the projection is evaluated
on $D_k$.

For the $S_1$ calculations reported here, the fixed ground state has $A_g$
symmetry, whereas the target states have $B_{2g}$ or $A_u$ symmetry. The
fixed-ground projection therefore does not change the residual in these
calculations. The projection is required in a general excited-state
calculation when the fixed ground state and the target state have the same
spatial symmetry.

\subsubsection{One-particle density matrix of the residual}
For a nonzero residual norm, define
\begin{equation}
 \nu_k=\sqrt{\braket{r_k}{r_k}},\qquad
 \ket{\bar r_k}=\frac{\ket{r_k}}{\nu_k}.
 \label{eq:si-resno-normalization}
\end{equation}
With $r_{k,A}=\braket{A}{r_k}$ on the support $D_k$, the spin-resolved 1-RDM of
the normalized residual is
\begin{equation}
 \gamma_{k,pq}^{\sigma,(R)}
 =\frac{\sum_{A,J\in D_k}r_{k,A}^*r_{k,J}
 \bra{A}\hat a^\dagger_{p\sigma}\hat a_{q\sigma}\ket{J}}
 {\nu_k^2}.
 \label{eq:si-resno-rdm-explicit}
\end{equation}
The residual 1-RDM is evaluated in the same way as the 1-RDM of the initial selected-CI wavefunction,
using the sign convention in Eq.~\eqref{eq:si-onebody-sign}. Its diagonal elements give the
orbital occupations, while its off-diagonal elements contain contributions from pairs of residual
determinants that differ by one spin-orbital substitution. For the normalized residual, the trace of
the spin-resolved 1-RDM is $N_\sigma$, and its natural occupation numbers lie between 0 and 1.
The RES-NO orbitals and frame are
\begin{equation}
 \gamma_k^{\sigma,(R)}R_{k,\sigma}^{(R)}
 =R_{k,\sigma}^{(R)}\Lambda_{k,\sigma}^{(R)},
 \qquad (R_{k,\sigma}^{(R)})^\dagger R_{k,\sigma}^{(R)}=I,
 \label{eq:si-resno-eigenproblem}
\end{equation}
\begin{equation}
 F_k^{(R)}=(R_{k,\alpha}^{(R)},R_{k,\beta}^{(R)}).
 \label{eq:si-resno-frame}
\end{equation}
The superscript $(R)$ denotes RES-NO. Hermiticity, the trace, and the
symmetry blocks are treated as for SCI-NO.
The residual is used only to construct the RES-NO orbitals. The external determinants
appearing in the residual are not added directly to the final selected-CI calculation.

If the residual is numerically zero, the initial selected-CI wavefunction has no detectable coupling to determinants outside its current space within the target symmetry. In this case, no residual 1-RDM can be constructed and RES-NO orbitals cannot be defined. Another frame, such as SCI-NO, can then be used instead.

\subsubsection{Selection of residual determinants for larger active spaces}
\label{si:resno-screening}

The calculations reported in this work use the complete residual. For larger
active spaces, the external determinant space may become too large, and the
residual can be approximated by selecting its most important determinants.
The external determinants $C_k$ are generated as in Sec.~\ref{si:resno-explicit}, and the retained candidates are
\begin{equation}
 C_{k,\epsilon}
 =
 \left\{
 A\in C_k
 \middle|
 \max_{I\in B}\left|H_{AI}c_I^{(k)}\right|>\epsilon
 \right\},
 \qquad
 \epsilon>0.
 \label{eq:si-resno-screen}
\end{equation}
The threshold is used only to select candidates. After the candidate set has
been determined, the residual coefficient of each retained determinant is
recalculated as
\begin{equation}
 u_{k,A}
 =
 \sum_{I\in B}H_{AI}c_I^{(k)},
 \qquad
 A\in C_{k,\epsilon}.
 \label{eq:si-resno-screened-coefficients}
\end{equation}
Thus, all contributions to a retained determinant are included, even when some
individual contributions are smaller than $\epsilon$. However, a determinant
can be excluded when all individual contributions are smaller than $\epsilon$,
even if their sum would be appreciable.

For an excited-state calculation, the fixed-ground projection is applied when
required. The resulting determinant set and residual coefficients are denoted
by $D_{k,\epsilon}$ and $r_{k,A}$. The $L$ determinants with the largest
squared residual coefficients are retained,
\begin{equation}
 \mathcal T_L
 =
 \operatorname{Top}_L
 \left\{
 |r_{k,A}|^2
 \middle|
 A\in D_{k,\epsilon}
 \right\}.
 \label{eq:si-resno-retained-set}
\end{equation}
The normalized truncated residual is
\begin{equation}
 \ket{\widetilde r_k}
 =
 \frac{\sum_{A\in\mathcal T_L}r_{k,A}\ket{A}}
 {\sqrt{\sum_{A\in\mathcal T_L}|r_{k,A}|^2}}.
 \label{eq:si-resno-retained-state}
\end{equation}
Its spin-resolved 1-RDM is diagonalized to obtain the approximate RES-NO
orbitals.

The fraction of the residual weight retained after selecting the largest $L$
contributions is
\begin{equation}
 f_{k,L}
 =
 \frac{\sum_{A\in\mathcal T_L}|r_{k,A}|^2}
 {\sum_{A\in D_{k,\epsilon}}|r_{k,A}|^2}.
 \label{eq:si-resno-retained-fraction}
\end{equation}
This quantity measures only the loss caused by the choice of $L$. It does not
include contributions removed by the threshold $\epsilon$.

Both $\epsilon$ and $L$ can affect the residual 1-RDM and the resulting RES-NO
orbitals. Their values should therefore be varied independently until the
orbitals and selected-CI energies are stable. Restricting the residual to $L$
determinants reduces the cost of constructing its 1-RDM, but it does not remove
the preceding cost of generating the candidate determinants and evaluating
their residual coefficients. The retained external determinants are used only
to construct the RES-NO orbitals and are not added directly to the final
selected-CI space.

\section{Classical candidate generation without full determinant enumeration}
\label{si:candidate-backtransform}
This section describes how canonical candidates are generated from the
rotated-frame measurements without evaluating the RSU score over the full
determinant space. Section~\ref{si:candidate-direct-sampling} summarizes the
procedure tested on the FCI-prepared N$_2$ ground-state measurements, and
Sec.~\ref{si:candidate-exact-closure} gives its theoretical basis and the
symmetry-aware variant used for the GQE-prepared ground states. In both
cases, the RSU score is evaluated only for an explicitly generated finite
candidate set, and no additional quantum shots are required.

\subsection{Finite candidate generation by direct back-transformation sampling}
\label{si:candidate-direct-sampling}

For an observed record $i=(k,F,d)$ with count $n_i$, let $W=\sum_i n_i$ and
$P_i(x)=|A^{(F)}_{d,x}|^2$ (Eq.~\eqref{eq:si-candidate-component}). A record
is chosen with probability $n_i/W$, and a canonical determinant is drawn from
$P_i(x)$. The generated candidates therefore follow the distribution
\begin{equation}
 q_x=\sum_i \frac{n_i}{W}P_i(x)=\frac{\sigma_x}{W},
 \label{eq:si-candidate-mixture}
\end{equation}
where $\sigma_x$ is the RSU score of Eq.~\eqref{eq:si-candidate-exact-score}.
Each draw uses the sequential determinantal sampler described in
Sec.~\ref{si:candidate-exact-closure}, applied separately to the two spins.
It stores only orbital-sized matrices, not a many-electron statevector or a
table over the full determinant space $\mathcal X$, and it preserves
$N_\alpha$ and $N_\beta$. Determinants observed in the identity frame are
included directly.

The tested procedure also adds pivoted-QR seed determinants obtained from the
observed rotated determinants, together with all same-spin single
substitutions of these seeds. All candidates are generated from the observed
rotated determinants, not from FCI coefficients. After duplicates are
removed, the RSU score
\begin{equation}
 \sigma_x=\sum_i n_iP_i(x),\qquad x\in\mathcal C,
 \label{eq:si-candidate-exact-score}
\end{equation}
is evaluated only on the generated set $\mathcal C$. Every observed record
contributes to the score of each candidate. Spin completion, symmetry
selection, and selected-CI diagonalization are then applied as in the main
text. The classical cost of this procedure is summarized at the end of
Sec.~\ref{si:candidate-exact-closure}.

\subsection{Candidate generation from the Slater expansion of each observed determinant}
\label{si:candidate-exact-closure}

The RSU score of Eq.~\eqref{eq:si-candidate-exact-score} is defined for every
canonical determinant $x$ in the full-CI determinant space $\mathcal X$, which
contains all determinants with $N_\alpha$ $\alpha$ electrons and $N_\beta$
$\beta$ electrons in the $n_{\mathrm{orb}}$ active orbitals. Because the size of
$\mathcal X$ grows combinatorially with the active space, we describe here how
canonical determinants with large scores are found without evaluating the
score over $\mathcal X$.

As in Sec.~\ref{si:candidate-direct-sampling}, an observed record $i=(k,F,d)$
denotes the rotated determinant with occupation string $d$ obtained by
measuring electronic state $k$ in orbital frame $F$; it was observed $n_i$
times, and $W=\sum_i n_i$. For a canonical determinant $x$,
\begin{equation}
 P_i(x)=|\det R_\alpha[x_\alpha,d_\alpha]|^2|\det R_\beta[x_\beta,d_\beta]|^2
 =|A^{(F)}_{d,x}|^2 .
 \label{eq:si-candidate-component}
\end{equation}
Here $R_\sigma$ $(\sigma=\alpha,\beta)$ is the
$n_{\mathrm{orb}}\times n_{\mathrm{orb}}$ orbital-rotation matrix of frame $F$
in Eq.~\eqref{eq:si-frame-orbitals}. Its $p$th column contains the
coefficients of the $p$th rotated orbital in the canonical molecular orbitals.
The frame label is omitted from $R_\sigma$ for brevity. The sets $d_\sigma$ and
$x_\sigma$ list the $N_\sigma$ occupied orbitals of spin $\sigma$ in $d$ and in
$x$. The $N_\sigma\times N_\sigma$ submatrix $R_\sigma[x_\sigma,d_\sigma]$,
formed by the rows $x_\sigma$ and the columns $d_\sigma$, is the overlap matrix
between the occupied canonical orbitals of $x$ and the occupied rotated
orbitals of $d$. Its determinant is the overlap of the two single-spin
determinants, and the product over both spins is the expansion coefficient
$A^{(F)}_{d,x}=\braket{x}{\varphi_{F,d}}$ of the main text. Thus $P_i(x)$ is the
weight of the canonical determinant $x$ in the observed rotated determinant.
It differs from the probability $P_{k,F}(d)$ of observing $d$ in the
measurement.

Because each rotated determinant is normalized, $\sum_{x\in\mathcal X}P_i(x)=1$.
The normalized score is therefore an average of these distributions weighted by
the measured counts,
\begin{equation}
 \frac{\sigma_x}{W}=\sum_i\frac{n_i}{W}P_i(x).
 \label{eq:si-candidate-score-mixture}
\end{equation}
Determinants with large scores can therefore be found by drawing canonical
determinants from $P_i$ for the observed records and collecting the
determinants that appear. In the identity frame, $R_\alpha=R_\beta=I$ and
$P_i(x)=\delta_{x,d}$, so only observed determinants have nonzero scores and
the procedure reduces to conventional QSCI selection.

A canonical determinant is sampled from $P_i$ by using the one-particle
density matrix of the rotated determinant, which for a single determinant is
a projector. For spin $\sigma$, let $V=R_\sigma[:,d_\sigma]$ be the
$n_{\mathrm{orb}}\times N_\sigma$ matrix of the occupied rotated orbitals of $d$,
and let
\begin{equation}
 K=VV^{\mathrm T}
 \label{eq:si-candidate-density}
\end{equation}
be the corresponding projector in the canonical orbitals. Its diagonal element
$K_{pp}$ is the occupation of canonical orbital $p$, and
$\operatorname{Tr}K=N_\sigma$. A first orbital $p$ is chosen with probability
$K_{pp}/N_\sigma$. Given that $p$ is occupied, the projector onto the
remaining $N_\sigma-1$ electrons is
\begin{equation}
 K'=K-\frac{Ke_pe_p^{\mathrm T}K}{K_{pp}},
 \label{eq:si-candidate-conditional-density}
\end{equation}
where $e_p$ is the unit vector of orbital $p$. Since $K'_{pp}=0$, the same
orbital cannot be chosen twice. The next orbital is chosen with probability
$K'_{qq}/(N_\sigma-1)$, and the step is repeated until $N_\sigma$ orbitals have
been chosen. For any ordering of a given set $x_\sigma$, the product of these
conditional probabilities equals $|\det R_\sigma[x_\sigma,d_\sigma]|^2/N_\sigma!$.
Summing over the $N_\sigma!$ orderings shows that the occupied set $x_\sigma$ is
obtained with probability $|\det R_\sigma[x_\sigma,d_\sigma]|^2$. The $\alpha$
and $\beta$ strings are sampled independently and combined into $x$, so that
$x$ is drawn from $P_i(x)$ and preserves $N_\alpha$ and $N_\beta$. Only
matrices of orbital dimension are used.

These samples are classical Monte Carlo samples and are distinct from the
quantum measurement shots. A total of $S$ samples is used, and each record $i$
is expanded at least $Sn_i/W$ times. Every canonical determinant obtained at
least once is placed in the candidate set $\mathcal C$. The sampled frequencies
are not used as scores. Instead, the score of each candidate is computed
exactly from all observed records,
\begin{equation}
 \sigma_x=\sum_i n_iP_i(x),\qquad x\in\mathcal C .
 \label{eq:si-candidate-rescoring}
\end{equation}
The ranking of the generated candidates therefore introduces no Monte Carlo
error beyond the statistical error of the measured counts. The only
approximation is that a determinant that is never sampled is not considered.
Its probability of being missed decreases exponentially with its score,
\begin{equation}
 \Pr(x\notin\mathcal C)\le\exp(-S\sigma_x/W),
 \label{eq:si-candidate-miss-probability}
\end{equation}
so that determinants with large scores are found with high probability.

Determinants near the selection threshold can still be missed. The candidate
set is therefore enlarged by single substitutions. Let $B$ be the maximum
number of determinants retained for the selected-CI calculation, let
$\mathcal Y^{(r)}$ be the $B$ candidates with the largest positive scores at
iteration $r$, and let $\mathcal N_1(x)$ be the determinants obtained from $x$ by moving
one electron to an unoccupied orbital of the same spin. Starting from
$\mathcal C^{(0)}=\mathcal C$, we set
\begin{equation}
 \mathcal C^{(r+1)}=\mathcal C^{(r)}\cup\bigcup_{x\in\mathcal Y^{(r)}}\mathcal N_1(x),
 \label{eq:si-candidate-closure-update}
\end{equation}
and evaluate Eq.~\eqref{eq:si-candidate-rescoring} only for the new
determinants. The iteration stops when no new neighbor enters the retained
set. At this point no single substitution of a retained determinant has a
larger score than the smallest retained score. This is a local condition and
does not guarantee that the retained set equals the exact top-$B$ set over
$\mathcal X$. Monte Carlo sampling reaches determinants far from the current
set, whereas the substitution step fills in neighbors near the threshold.

If the frame does not mix orbitals of different irreducible representations,
$P_i(x)$ is nonzero only for canonical determinants $x$ that have the same
spin-resolved occupation of each irreducible-representation block as $d$.
Determinants of other spatial symmetries may still appear as candidates, for
example from observed records with nontarget symmetry or from frames that mix
irreducible representations, but they are excluded from the final
selected-CI space.

\paragraph{Symmetry-aware deterministic candidate generation.}
For the GQE-prepared ground-state calculations, we use a deterministic,
symmetry-aware variant of the procedure described above. The initial candidate
set contains the determinants observed in the identity frame and the
pivoted-QR seed determinants (Sec.~\ref{si:candidate-direct-sampling})
obtained from the observed rotated determinants; no Monte Carlo draws are
used. We refer to this construction of the initial set as multideterminant
unrotation seeding (MDUS).

Because the orbital rotations are block diagonal with respect to the spatial
irreducible representations, each observed rotated determinant expands only
into canonical determinants with the same spin-resolved occupation of every
irreducible-representation block. Accordingly, $\mathcal N_1(x)$ in
Eq.~\eqref{eq:si-candidate-closure-update} is restricted to same-spin single
substitutions that preserve these block occupations, and the score of
Eq.~\eqref{eq:si-candidate-rescoring} is evaluated only for newly generated
candidates. We refer to this symmetry-preserving expansion as
single-excitation closure (SXC); it stops when no new determinant enters the
retained set. The procedure requires the irreducible representations of the
orbitals and the target symmetry sector, but neither FCI coefficients nor an
FCI wavefunction. When the target symmetry is not known in advance, each
relevant symmetry sector must be treated separately.

In the calculations reported here, the retained-set size was
$B_{\mathrm{closure}}=2000$, the number of closure iterations was limited to
100, and the candidate set was limited to 100,000 determinants. MDUS+SXC does
not enumerate the complete fixed-electron-number determinant space. Without
these limits, repeated closure could in principle expand over a large fraction
of the target symmetry sector. The limits therefore bound the classical cost
but do not guarantee recovery of the exact top-$B$ determinants over
$\mathcal X$.

The classical cost is set by the number of Monte Carlo samples $S$, the number
of distinct observed records $N_{\mathrm{rec}}$, the number of evaluated
candidates $N_{\mathrm{cand}}$, the cap $B$, and the orbital and electron
numbers. Each draw of the sequential sampler requires
$O(n_{\mathrm{orb}}^2)$ memory and $O(N_\sigma n_{\mathrm{orb}}^2)$ operations per
spin. Scoring requires at most $N_{\mathrm{rec}}N_{\mathrm{cand}}$ evaluations
of Eq.~\eqref{eq:si-candidate-component}, each involving determinants of
$N_\sigma\times N_\sigma$ matrices. One substitution step generates at most
$B\sum_\sigma N_\sigma(n_{\mathrm{orb}}-N_\sigma)$ new determinants. None of
these quantities depends explicitly on the size of $\mathcal X$. These bounds
describe the cost of the post-processing for given numbers of samples and
candidates; they do not guarantee energy accuracy for arbitrary active spaces.

\section{Canonical determinant recovery for the GQE-prepared ground state}
\label{sec:si-s0-gqe-support}
\begin{figure}[ht]
\centering
\includegraphics[width=\linewidth]{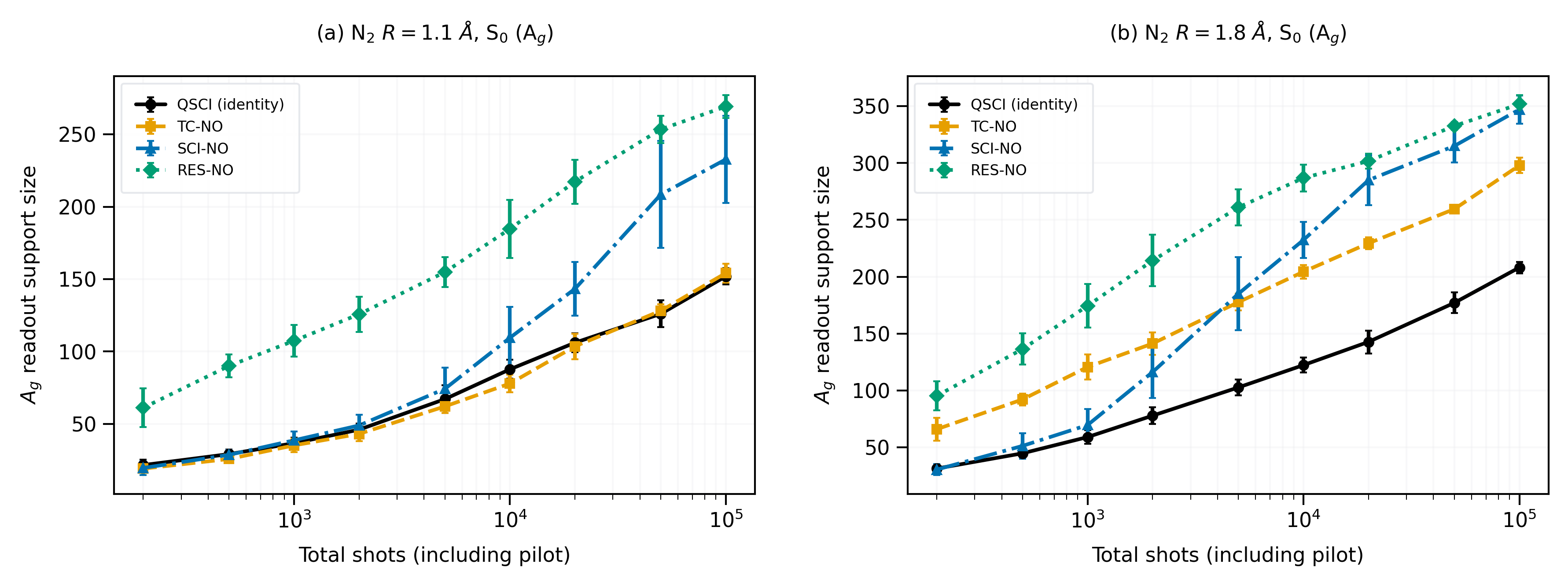}
\caption{Mean number of canonical $A_g$ determinants in the final selected-CI
space versus the total number of shots for the generalized-pool GQE-prepared
$S_0$ states, with at most 2000 determinants in the selected-CI space.
Left: $R=1.1$~\AA; right: $R=1.8$~\AA. Symbols are means over eight
measurement seeds after averaging the three independently trained GQE
references within each measurement seed; error bars are sample standard
deviations.}
\label{fig:si-s0-gqe-support}
\end{figure}

Figure~\ref{fig:si-s0-gqe-support} shows the number of canonical $A_g$
determinants in the final selected-CI spaces for the GQE-prepared ground
states of main-text Fig.~\ref{fig:rsu-gqe-s0}. The plotted values are the
numbers of determinants actually retained, not the maximum allowed number.

At $R=1.1$~\AA\ and 200 shots, RES-NO retains 61 canonical $A_g$ determinants,
compared with 22 for conventional QSCI. At $10^5$ shots, RES-NO retains 269
and SCI-NO 233. The TC-NO rotation constructed from the GQE circuit is very
small at this geometry, so the TC-NO results remain close to those of
conventional QSCI. At $R=1.8$~\AA, RES-NO retains 95 determinants at 200 shots
and 352 at $10^5$ shots, compared with 31 and 208 for QSCI. At $10^5$ shots,
SCI-NO retains 347 and TC-NO 298.

Thus, SCI-NO and RES-NO include more canonical determinants in the
selected-CI space than conventional QSCI at the same total number of shots.
This broader determinant recovery helps explain the lower ground-state
energies obtained with SCI-NO and RES-NO in the main text. Because the
methods retain different numbers of determinants at the same number of
shots, the figure does not compare the methods at a fixed CI dimension.

% ----------------------------------------------------------------------

\section{Interference in RSU proposal scores and amplitude-based scoring}
\label{si:rsu-interference}
\subsection{Interference terms omitted from the RSU proposal score}
For a normalized prepared state $\ket{\Psi_k}$ and an orbital frame $F$, we
define the amplitudes in the rotated and canonical determinant bases as
\begin{equation}
 b_{k,F,d}=\braket{\varphi_{F,d}}{\Psi_k},\qquad
 c_{k,x}=\braket{x}{\Psi_k}
 =\sum_{d}A^{(F)}_{d,x}b_{k,F,d},
 \label{eq:si-rsu-amplitudes}
\end{equation}
where the sum runs over all rotated determinants and
$A^{(F)}_{d,x}=\braket{x}{\varphi_{F,d}}$ as in the main text. With
$P_{k,F}(d)=|b_{k,F,d}|^2$, the probability of canonical determinant $x$ is
\begin{equation}
 |c_{k,x}|^2=q_{k,F,x}+J_{k,F,x},\qquad
 q_{k,F,x}=\sum_d |A^{(F)}_{d,x}|^2P_{k,F}(d),
 \label{eq:si-rsu-weight}
\end{equation}
where
\begin{equation}
 J_{k,F,x}=2\operatorname{Re}\sum_{d<e}
 A^{(F)}_{d,x}(A^{(F)}_{e,x})^*
 b_{k,F,d}b_{k,F,e}^*
 \label{eq:si-rsu-interference}
\end{equation}
is the interference between different rotated determinants; the restriction
$d<e$ counts each pair once. This interference arises when the canonical
amplitude is reconstructed from the rotated amplitudes. It is different from
the interference between canonical determinants that is already contained in
the measured probability $P_{k,F}(d)$ through
$b_{k,F,d}=\sum_x(A^{(F)}_{d,x})^*c_{k,x}$.

If $N_{k,F}$ shots are measured for state $k$ in frame $F$, the expected
count of $d$ is $N_{k,F}P_{k,F}(d)$. The expectation value of the RSU score in
main-text Eq.~\eqref{eq:rsu-proposal-score} is therefore
\begin{equation}
 \mathbb E[\sigma_x]=\sum_{k,F}N_{k,F}q_{k,F,x},
 \label{eq:si-rsu-score-expectation}
\end{equation}
which in general differs from $\sum_{k,F}N_{k,F}|c_{k,x}|^2$. Equivalently,
$q_{k,F,x}$ is the canonical diagonal element of the dephased state
\begin{equation}
 \rho_{k,F}=\sum_d P_{k,F}(d)
 \ket{\varphi_{F,d}}\bra{\varphi_{F,d}},
 \label{eq:si-rsu-dephasing}
\end{equation}
in which the coherences between rotated determinants are removed. The RSU
score is therefore an incoherent measure used to find determinants, not an
estimate of a CI coefficient or of the canonical probability $|c_{k,x}|^2$.
Because the frame is unitary, $\sum_xq_{k,F,x}=\sum_x|c_{k,x}|^2=1$ and hence
$\sum_xJ_{k,F,x}=0$; the interference redistributes weight among canonical
determinants without changing its total. It vanishes in the identity frame
and when the prepared state is a single rotated determinant, but not in
general. Including it requires the relative signs of the rotated amplitudes,
which cannot be obtained by taking square roots of the counts.

\subsection{Estimation of canonical determinant amplitudes from~measurement counts}
Measurements in different orbital frames give probabilities but not the
relative signs of the canonical amplitudes. We therefore tested a numerical
procedure that fits one common set of real canonical amplitudes to the
histograms measured in all frames.

Let $z_x$ be a real trial amplitude of canonical determinant $x$ in a
restricted set $\mathcal C$. The probability of observing rotated determinant
$d$ in frame $F$ is modeled as
\begin{equation}
 p_{F,d}(z)=\left|\sum_{x\in\mathcal C}
 (A^{(F)}_{d,x})^*z_x\right|^2.
 \label{eq:si-rsu-coherent-model}
\end{equation}
Because the amplitudes are summed before the absolute square is taken,
Eq.~\eqref{eq:si-rsu-coherent-model} includes the interference between
canonical determinants. The trial amplitudes are obtained by minimizing
\begin{equation}
 \mathcal L(z)=N\sum_{x\in\mathcal C}|z_x|^2
 -\sum_{F,d:n_{F,d}>0}n_{F,d}
 \log(p_{F,d}(z)+\eta),\qquad
 N=\sum_F N_F,
 \label{eq:si-rsu-coherent-loss}
\end{equation}
where $n_{F,d}$ is the count of $d$ in frame $F$, $N_F$ is the number of shots
in frame $F$, and $\eta=10^{-14}$ prevents evaluation of the logarithm at
zero. Because each frame is unitary, $\sum_dp_{F,d}(z)=\sum_x|z_x|^2$, so the
first term equals $\sum_FN_F\sum_dp_{F,d}(z)$ and $\mathcal L$ is the
negative log-likelihood of the counts under independent Poisson statistics.
The fitted amplitudes are normalized as
$\widehat c=z/\sqrt{\sum_x|z_x|^2}$, and determinant $x$ is scored by
\begin{equation}
 s_x=N|\widehat c_x|^2.
 \label{eq:si-rsu-coherent-score}
\end{equation}
This score is used only to select canonical determinants; the CI
coefficients are then obtained by diagonalizing the Hamiltonian in the
selected space. The fitted amplitudes are approximate because the counts
contain finite-shot errors, only a restricted set of canonical determinants is
included, and the minimization can depend on the initial amplitudes.

\subsection{Results with amplitudes estimated from~measurement counts}
\label{si:rsu-coherent-results}
As a preliminary test, we reanalyzed previously obtained measurements of the
FCI-prepared $S_1$ states of N$_2$/STO-3G $(10e,8o)$: $S_1(B_{2g})$ at
$R=1.1$~\AA\ and $S_1(A_u)$ at $R=1.8$~\AA. The same histograms were scored
either with the frequency-based RSU score or with the score of
Eq.~\eqref{eq:si-rsu-coherent-score}. The calculation conditions are listed
in Table~\ref{tab:si-coherent-setup}, and the results are summarized in
Table~\ref{tab:si-coherent-results}.
\begin{table}[ht]
\centering
\caption{Conditions of the preliminary test with amplitudes estimated from
measurement counts.}
\label{tab:si-coherent-setup}
\begin{tabular}{ll}
\toprule
Sampled states & exact FCI $S_1$ wavefunctions \\
Total shots & nine values from 200 to 100,000 \\
Measurement seeds & 8 \\
Determinants in each amplitude fit & at most 512 \\
Initial amplitude sets & 4 (16 and 64 in the stability check) \\
Iterations per minimization & at most 200 \\
Determinants in the final selected-CI space & at most 2000, after spin completion \\
\bottomrule
\end{tabular}
\end{table}
\begin{table}[ht]
\centering
\caption{Change in the mean absolute $S_1$ energy error when the
frequency-based score is replaced by the score from fitted amplitudes
(positive values indicate larger errors).}
\label{tab:si-coherent-results}
\begin{tabular}{lll}
\toprule
Frame & $R=1.1$~\AA & $R=1.8$~\AA \\
\midrule
QSCI   & identical & identical \\
SCI-NO & nearly unchanged; no decrease & nearly unchanged; no decrease \\
RES-NO & nearly unchanged; no decrease & nearly unchanged; no decrease \\
\bottomrule
\end{tabular}
\end{table}

QSCI is unaffected because, in the identity frame, the fitted $|z_x|^2$ are
proportional to the observed counts and give the same ranking. For SCI-NO
and RES-NO, the fitted amplitudes changed the selected determinants in some
calculations, but the mean absolute $S_1$ error was nearly unchanged and did
not decrease in any case.

Because the amplitudes are obtained by a nonlinear fit, the lack of
improvement could originate from the fit itself. We therefore repeated the fit
from more initial guesses. The energy differences from the frequency-based
score kept the same sign, so the result is not an artifact of the particular
initial guesses. However, determinants with fitted amplitudes close to zero
were selected in some fits and not in others, so full convergence of the fit
and the stability of the selected determinants could not be confirmed.

These results do not show that the interference terms are harmful. They show
only that a single nonlinear fit of the amplitudes to finite-shot counts does
not yet extract useful interference information. Whether an improved
treatment, for example of the fitting procedure or of determinants with small
fitted amplitudes, can increase the sampling efficiency of RSU remains a
subject for future work. The main-text results use the frequency-based score
and are not affected by this test.

\section{Wavefunction refinement and its combination with RSU}
\label{si:rsu-refinement}
\subsection{Wavefunction refinement}
\label{si:wavefunction-refinement}
Wavefunction refinement follows the local and global workflow of
Ref.~\cite{kemmoku2026}. It is an optional classical step applied after
canonical QSCI or RSU selection. A GQE policy can provide several circuits.
If only the selected-CI space of one circuit is kept, determinants found with
the other circuits are lost. Wavefunction refinement combines the sparse CI
wavefunctions obtained from different circuits and iterations and keeps the
determinants that are most important in the combined wavefunction, within the
same determinant cap $B$.

Let $\ket{\psi^{(1)}},\ldots,\ket{\psi^{(M)}}$ be normalized selected-CI
wavefunctions in the common canonical determinant basis,
\begin{equation}
 \ket{\psi^{(j)}}=\sum_{x\in\mathcal B^{(j)}}a_x^{(j)}\ket{x},
 \qquad
 \mathcal U=\bigcup_{j=1}^{M}\mathcal B^{(j)},
 \label{eq:si-refinement-inputs}
\end{equation}
where $\mathcal B^{(j)}$ is the selected determinant space of input $j$ and
$a_x^{(j)}=0$ for $x\notin\mathcal B^{(j)}$. The input wavefunctions are
expanded in orthonormal canonical determinants, but they are generally not
orthogonal to each other. Their Hamiltonian and overlap matrices are
\begin{equation}
 H^{\mathrm{ref}}_{ij}=\bra{\psi^{(i)}}\hat H\ket{\psi^{(j)}}
 =\sum_{x\in\mathcal B^{(i)}}\sum_{y\in\mathcal B^{(j)}}
 a_x^{(i)}H_{xy}a_y^{(j)},
 \qquad
 S^{\mathrm{ref}}_{ij}=\braket{\psi^{(i)}}{\psi^{(j)}}
 =\sum_{x\in\mathcal U}a_x^{(i)}a_x^{(j)},
 \label{eq:si-refinement-matrices}
\end{equation}
for real coefficients. The combined wavefunctions are obtained from the
generalized eigenvalue problem
\begin{equation}
 H^{\mathrm{ref}}\mathbf c_k=E_k^{\mathrm{ref}}S^{\mathrm{ref}}\mathbf c_k .
 \label{eq:si-refinement-gevp}
\end{equation}
Nearly linearly dependent inputs make $S^{\mathrm{ref}}$ ill conditioned.
We therefore use canonical orthogonalization. With
$S^{\mathrm{ref}}=V s V^{T}$, where $s=\operatorname{diag}(s_1,\ldots,s_M)$,
the eigenvectors with $s_n<10^{-10}$ are removed, and
Eq.~\eqref{eq:si-refinement-gevp} is solved as an ordinary eigenvalue problem
in the remaining orthonormal combinations.

The combined wavefunction for root $k$ has the canonical coefficients
\begin{equation}
 \bar a_{k,x}=\sum_{j=1}^{M}c_{k,j}a_x^{(j)},
 \qquad x\in\mathcal U .
 \label{eq:si-refinement-amplitudes}
\end{equation}
For a ground-state calculation, the $B$ determinants with the largest
$|\bar a_{0,x}|^2$ are retained. For several target roots, a state-averaged
weight can be used instead,
\begin{equation}
 \bar p_x=\sum_{k\in\mathcal K}\omega_k|\bar a_{k,x}|^2,
 \qquad
 \mathcal B_{\mathrm{ref}}
 =\operatorname{Top}_{B}\left\{\bar p_x \middle| x\in\mathcal U\right\},
 \label{eq:si-refinement-recompression}
\end{equation}
where $\mathcal K$ is the set of target roots and $\omega_k\ge0$ their
weights. The Hamiltonian is then diagonalized again in the retained canonical
determinants $\mathcal B_{\mathrm{ref}}$, which gives new CI coefficients. We
denote this sequence of mixing, retention, and rediagonalization by
\begin{equation}
 \ket{\psi'}=\mathcal R\left(\ket{\psi^{(1)}},\ldots,\ket{\psi^{(M)}}\right).
 \label{eq:si-refinement-operation}
\end{equation}
The generalized eigenvalue problem in Eq.~\eqref{eq:si-refinement-gevp} is
used only to combine the input wavefunctions. The final selected-CI problem is
solved in orthonormal canonical determinants.

The operation $\mathcal R$ is applied at two levels. Let
$\ket{\psi_i^{(m)}}$ be the selected-CI wavefunction obtained from circuit $m$
at iteration $i$, with $m=1,\ldots,M$. Local refinement combines the circuits
of the same iteration,
\begin{equation}
 \ket{\psi_i^{L}}
 =\mathcal R\left(\ket{\psi_i^{(1)}},\ldots,\ket{\psi_i^{(M)}}\right),
 \label{eq:si-refinement-local}
\end{equation}
and global refinement carries the result to the next iteration,
\begin{equation}
 \ket{\psi_1^{G}}=\ket{\psi_1^{L}},
 \qquad
 \ket{\psi_i^{G}}=\mathcal R\left(\ket{\psi_{i-1}^{G}},\ket{\psi_i^{L}}\right),
 \qquad i\ge2 .
 \label{eq:si-refinement-global}
\end{equation}
For QSCI, the inputs $\ket{\psi_i^{(m)}}$ are obtained from canonical-basis
measurements. For RSU, they are obtained from rotated-frame measurements
followed by canonical selection.

Wavefunction refinement uses only CI coefficients and Hamiltonian matrix
elements that are already available, so it requires no additional quantum
measurements. It does require additional classical Hamiltonian construction and
diagonalization. It cannot add determinants outside the union $\mathcal U$ of
its inputs. Because the retained determinants are chosen by CI weight and not
by energy, a single refinement step is not guaranteed to lower the energy.

\subsection{Combined use with RSU for the ground state}
\label{si:rsu-refinement-combined}
\label{si:frame-refinement-separation}
RSU and wavefunction refinement act on different steps. RSU changes the
orbital frame used for measurement and thereby increases the number of
canonical determinants found from the same number of shots. Wavefunction
refinement chooses which of the found determinants are kept when the size of
the selected-CI space is limited. Both work with determinants in the same
canonical basis and can be applied one after the other.

We examined this combination for the FCI-prepared ground state of
N$_2$/STO-3G $(10e,8o)$ at $R=1.1$ and $1.8$~\AA\ ($A_g$), using the
measurements of main-text Fig.~\ref{fig:rsu-matched-n2}. No additional
measurements were made. Because wavefunction refinement combines CI
wavefunctions from several circuits, whereas this test has only one prepared
state, the measurements were divided into groups. For each method (QSCI,
SCI-NO, and RES-NO), total number of shots $N$, and measurement seed,
the $N$ shots were divided into four groups of $N/4$ shots, and each group was
treated as the measurement result of a separate circuit. The total number of
shots is therefore $N$ with and without refinement. The orbital frame was
kept fixed for all groups. A
selected-CI calculation was performed for each group. In the first
iteration, groups 1 and 2 were combined by local refinement
(Eq.~\eqref{eq:si-refinement-local}), and the result was carried over as the
global state. In the second iteration, groups 3 and 4 were combined by local
refinement, and the result was combined with the carried global state by
global refinement (Eq.~\eqref{eq:si-refinement-global}). Every selected-CI
space was limited to at most 170 determinants. The refined result was
compared with the ordinary selected-CI calculation that uses all $N$ shots
together under the same limit. Because the four groups come from the same
prepared state, this test examines how the determinants are chosen, not
whether new determinants are found.
Eight measurement seeds were used. Paired 95\% confidence intervals over the
seeds were evaluated for each bond length and number of shots, giving 18
conditions per method.

\begin{figure}[ht]
\centering
\includegraphics[width=\linewidth]{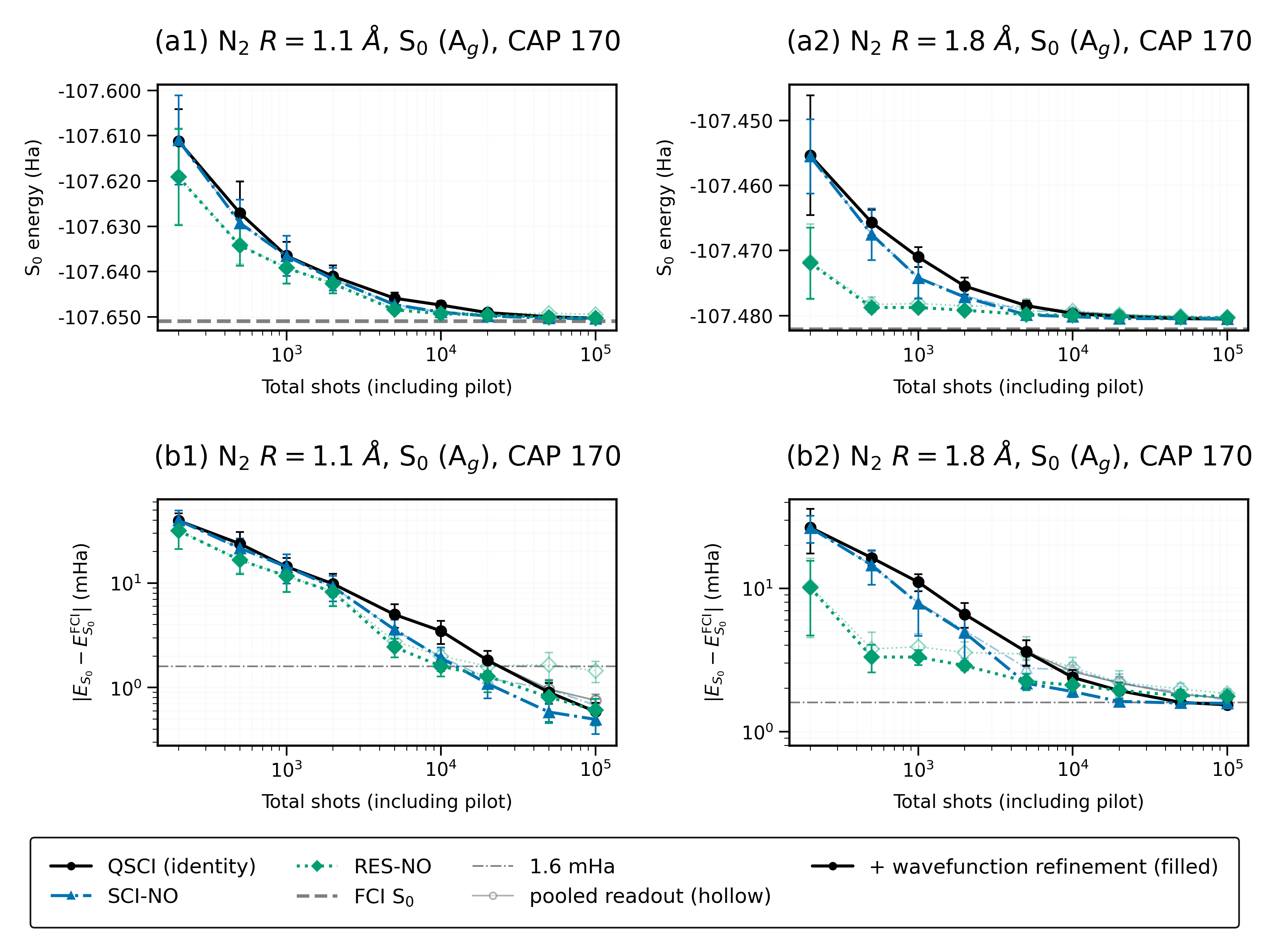}
\caption{(a) Energy and (b) absolute energy error of the FCI-prepared N$_2$
$S_0$ state versus the total number of shots, with at most 170 determinants in
the selected-CI space. Left: $R=1.1$~\AA; right: $R=1.8$~\AA. Colors, symbols,
and line styles identify the methods. Faint lines with open symbols show the
selected-CI calculation using all $N$ shots together, without refinement.
Bold lines with filled symbols show the same measurements after wavefunction
refinement of four groups of $N/4$ shots.
For SCI-NO and RES-NO, the total number of shots includes the initial
canonical-basis measurements used to construct the orbital frames, so all
methods are compared at the same total number of shots. Symbols are means over
eight measurement seeds, and error bars are sample standard deviations.
Dashed lines mark the FCI energies, and dash-dotted lines mark 1.6~mHa.}
\label{fig:si-refinement-cap170}
\end{figure}

Figure~\ref{fig:si-refinement-cap170} shows the results, and
Table~\ref{tab:si-refinement-errors} lists the errors at three numbers of
shots. For the same measured data, refinement generally lowered the mean
error of each method. The reduction was statistically resolved in 5, 6,
and 12 of the 18 conditions for QSCI, SCI-NO, and RES-NO, respectively, and
no condition showed a resolved increase in error. The largest gain was
obtained for RES-NO, presumably because its frequency-based ranking is least
suited to choosing determinants when the selected-CI space is limited.

The combination of a rotated frame and refinement gave the lowest mean error
in 17 of the 18 conditions. For example, at $R=1.8$~\AA\ and $N=500$, RES-NO
with refinement gave an error of $3.3$~mHa, compared with $16.4$~mHa for
ordinary QSCI. In 8 of these 17 conditions, the error of the combination was
lower than both that of the same rotated frame without refinement and that of
QSCI with refinement, with the paired confidence intervals excluding zero. For RES-NO, the gain from
refinement was larger than that for QSCI, with a statistically resolved
difference in 9 of the 18 conditions.
When the limit is raised from 170 to 2000 determinants, each method gives almost
the same results with and without refinement.

In conclusion, for the ground state with a finite number of shots and a limited number of determinants
in the selected-CI space, the combination of RSU and wavefunction refinement gives the most
accurate energies among the methods examined in nearly all conditions.
The two methods improve the energy in a
complementary way. RSU enlarges the set of candidate determinants obtained from
the same number of shots, and wavefunction refinement selects better among
these candidates.

These results have three limitations. First, refinement keeps the
determinants with the largest CI weights and does not minimize the energy
directly, so an individual refinement step occasionally raised the energy, by
at most $2.0$~mHa.
Second, the four groups of $N/4$ shots were all measured from the same FCI ground state.
Refinement could therefore only choose among the determinants that already appeared in these measurements.
Whether combining wavefunctions obtained from different GQE circuits or iterations also finds new determinants
was not examined here. Third, refinement requires additional classical calculations, namely a selected-CI
diagonalization for each group and further diagonalizations to combine the resulting wavefunctions.

\begin{table}[ht]
\centering
\caption{Mean absolute $S_0$ energy errors (mHa) without (w/o) and with (w/)
wavefunction refinement, taken from Fig.~\ref{fig:si-refinement-cap170}. The
selected-CI space contains at most 170 determinants. The lowest error in each
row is shown in bold.}
\label{tab:si-refinement-errors}
\begin{tabular}{llcccccc}
\toprule
 & & \multicolumn{2}{c}{QSCI} & \multicolumn{2}{c}{SCI-NO} & \multicolumn{2}{c}{RES-NO}\\
\cmidrule(lr){3-4}\cmidrule(lr){5-6}\cmidrule(lr){7-8}
$R$ (\AA) & Shots & w/o & w/ & w/o & w/ & w/o & w/ \\
\midrule
1.1 & 500 & 23.7 & 23.8 & 21.6 & 21.6 & 16.8 & \textbf{16.6} \\
 & 5000 & 5.01 & 5.01 & 3.60 & 3.59 & 2.78 & \textbf{2.45} \\
 & $10^5$ & 0.75 & 0.59 & 0.68 & \textbf{0.50} & 1.45 & 0.61 \\
\midrule
1.8 & 500 & 16.4 & 16.4 & 14.6 & 14.6 & 3.77 & \textbf{3.32} \\
 & 5000 & 3.63 & 3.61 & 2.78 & \textbf{2.18} & 3.48 & 2.24 \\
 & $10^5$ & 1.70 & \textbf{1.54} & 1.65 & 1.57 & 1.84 & 1.76 \\
\bottomrule
\end{tabular}
\end{table}

\section{Additional quantum gates for rotated-frame measurements}
\label{si:frame-circuit-cost}
Rotated-frame measurements require an additional orbital-rotation circuit
after state preparation. Table~\ref{tab:si-orbital-rotation-gates} lists
the gate counts for the $(10e,8o)$ active space. For a general real rotation
of $n$ spatial orbitals, at most $n(n-1)/2$ Givens rotations are needed per
spin. The RES-NO rotation examined here uses far fewer rotations.

The gate count depends on the assignment of spin orbitals to qubits and
the qubit connections. Placing each pair of spin orbitals to be mixed on
adjacent qubits avoids additional gates for fermionic parity. This requires
state preparation in the same spin-orbital ordering. With alternating
$\alpha$ and $\beta$ spin orbitals, additional gates account for fermionic
parity between separated orbitals. If two-qubit gates are restricted to
neighboring qubits in a one-dimensional array, fermionic SWAP gates are
also used to bring the required spin orbitals next to each other.

\begin{table}[ht]
\centering
\caption{Additional orbital-rotation gates for the $(10e,8o)$ active space.
State-preparation gates are excluded. The general upper bound assumes
adjacent spin orbitals and excludes gates for rearranging them. The RES-NO
counts refer to the N$_2$ circuits constructed here and have not been
minimized. The adjacent-pair depths assume parallel execution of rotations
on disjoint qubit pairs. A dash indicates a count not reported.}
\label{tab:si-orbital-rotation-gates}
\begin{tabular}{p{0.46\linewidth}rrrr}
\toprule
Orbital rotation and qubit arrangement & CNOT & One-qubit & Two-qubit & Total \\
& gates & gates & depth & depth \\
\midrule
General rotation (at most 56 Givens rotations) & 112 & 224 & --- & --- \\
\midrule
RES-NO (4 Givens rotations), with each orbital pair on adjacent qubits
& 8 & 16 & 2 & 5 \\
RES-NO, with alternating $\alpha$ and $\beta$ spin orbitals and direct
gates between any qubit pair & 64 & --- & 62 & --- \\
RES-NO, with gates restricted to neighboring qubits in a one-dimensional
array, including fermionic SWAP gates & 120 & 296 & --- & --- \\
\bottomrule
\end{tabular}
\end{table}

Errors in these additional gates can affect the measured outcomes and
determinant selection. The reduction in measurement cost observed without
gate errors must therefore be verified with noisy simulations or hardware
experiments that include the orbital-rotation circuit.

The SCI-NO and RES-NO rotation angles are determined from the initial
selected-CI calculation and kept fixed for all subsequent measurements
in the corresponding orbital frame.

RSU improves finite-shot determinant discovery at the cost of additional orbital-rotation gates.
Its practical benefit therefore depends on whether the reduction in measurement requirements outweighs
the additional circuit cost and gate errors.

\end{document}